\documentclass{aa}	     

\usepackage{epsfig}
\usepackage{graphicx}
\usepackage{epstopdf}
\usepackage{txfonts}
\usepackage{natbib}
\usepackage{color}
\usepackage{textcomp}
\usepackage[ ]{hyperref}    
\usepackage{amsmath}
\usepackage{amssymb}
\usepackage{gensymb}
\usepackage{tikz}
\usepackage{txfonts}
\usepackage{longtable}
\usepackage{array,multirow}

\newcommand{\Msun}{ {\rm M_{ \odot}}/h }
\newcommand{\Ha}{${\rm H{\alpha}}\,$}
\newcommand{\Hb}{${\rm H{\beta}}\,$}
\newcommand{\oii}{{\rm [\ion{O}{II}]}}
\newcommand{\oi}{{\rm [\ion{O}{I}]}}
\newcommand{\oiii}{{\rm [\ion{O}{III}]}}
\newcommand{\oiiiFd}{{\rm $\rm [\ion{O}{III}]_{5007}$}}
\newcommand{\usdss}{${ u\,}$}
\newcommand{\gsdss}{${ g\,}$}
\newcommand{\rsdss}{${ r\,}$}
\newcommand{\isdss}{${ i\,}$}
\newcommand{\zsdss}{${ z\,}$}
\newcommand{\jha}{${J0660}$}
\newcommand{\jOII}{${J0515}$}

\newcommand{\lgs}{\texttt{L-Galaxies} }

\definecolor{myorange}{RGB}{220,100,40}
\definecolor{myblue}{rgb}{0.0, 0.5, 0.69}

\begin{document}

\title{J-PLUS: Synthetic galaxy catalogues with emission lines for photometric surveys}
\author{David Izquierdo-Villalba$^{*1}$ \and Raul E. Angulo$^{2,3}$ \and Alvaro Orsi$^{1}$ \and  Guillaume Hurier$^{1}$ \and Gonzalo Vilella-Rojo$^{1}$ \and  Silvia Bonoli$^{2,3}$ \and Carlos López-Sanjuan$^{4}$ \and Jailson Alcaniz$^{5,6}$ \and Javier Cenarro$^{4}$ \and David Cristóbal-Hornillos$^{4}$ \and Renato Dupke$^{5}$ \and Alessandro Ederoclite$^{7}$ \and Carlos Hernández-Monteagudo$^{4}$ \and Antonio Marín-Franch$^{4}$ \and Mariano Moles$^{4}$ \and Claudia Mendes de Oliveira$^{7}$ \and Laerte Sodré Jr.$^{7}$ \and Jesús Varela$^{4}$ \and Héctor Vázquez Ramió$^{4}$}
\institute{$^{1}$ Centro de Estudios de Física del Cosmos de Aragón, Plaza San Juan 1, 44001 Teruel, Spain\\
$^{2}$ Donostia International Physics Centre (DIPC), Paseo Manuel de Lardizabal 4, 20018 Donostia-San Sebastian, Spain\\
$^{3}$ IKERBASQUE, Basque Foundation for Science, E-48013, Bilbao, Spain\\
$^{4}$ Centro de Estudios de Física del Cosmos de Aragón (CEFCA), Unidad Asociada al CSIC, Plaza San Juan 1, 44001 Teruel, Spain\\
$^{5}$ Observat\'{o}rio Nacional, CEP 20921-400, S\~{a}o Crist\'{o}v\~{a}o, Rio de Janeiro-RJ, Brazil.\\
$^{6}$ Departamento de Física, Universidade Federal do Rio Grande do Norte, 59072-970, Natal, RN, Brasil\\
$^{7}$ Institute of Astronomy, Geophysics and Atmospheric Sciences, University of S\~{a}o Paulo, CEP 05508-090, Cidade Universit\'{a}ria, S\~{a}o  Paulo-SP, Brazil. Brazil\\
\email{dizquierdo@cefca.es}}
\date{Received; accepted}
	 
\abstract  
{
We present a synthetic galaxy lightcone specially designed for narrow-band optical photometric surveys. To reduce time-discreteness effects, unlike previous works, we directly include the lightcone construction in the \texttt{L-Galaxies} semi-analytic model applied to the subhalo merger trees of the {\tt Millennium} simulation. Additionally, we add a model for the nebular emission in star-forming  regions, which is crucial for correctly predicting the narrow- and medium-band photometry of galaxies. Specifically, we consider, individually for each galaxy, the contribution of 9 different lines: $\rm Ly{\alpha}$ (1216\AA), \Hb (4861\AA), \Ha (6563\AA), {\oii} (3727\AA, 3729\AA), {\oiii} (4959\AA, 5007\AA), $\rm [\ion{Ne}{III}]$ (3870\AA), {\oi} (6300\AA), $\rm [\ion{N}{II}]$ (6548\AA, 6583\AA), and $\rm [\ion{S}{II}]$ (6717\AA, 6731\AA). We validate our lightcone by comparing galaxy number counts, angular clustering, and \Ha, \Hb, {\oii}, and {\oiiiFd} luminosity functions to a compilation of observations. As an application of our mock lightcones, we generated catalogues tailored for J-PLUS, a large optical galaxy survey featuring five broad-band and seven medium-band filters. We study the ability of the survey to correctly identify,  with a simple three-filter method, a population of emission-line galaxies at various redshifts. We show that the $4000\AA$ break in the spectral energy distribution of galaxies can be misidentified as line emission. However, all significant excess (larger than 0.4 magnitudes) can be correctly and unambiguously attributed to emission-line galaxies. Our catalogues are publicly available at \href{https://www.j-plus.es/ancillarydata/mock\_galaxy\_lightcone}{https://www.j-plus.es/ancillarydata/mock\_galaxy\_lightcone} to facilitate their use in interpreting narrow-band surveys and in quantifying the impact of line emission in broad-band photometry.\\ 
}
\titlerunning{Mock galaxy lightcones with emission lines for photometric surveys}
\authorrunning{Izquierdo-Villalba et al}
\maketitle
\newpage
\section{Introduction}

Optical surveys have been important in establishing our current understanding  of how galaxies form and evolve \citep{York2000, Gunn2006, Eisenstein2011,Driver2009, Grogin2011, Koekemoer2011, Sanchez2012, Dawson2013,SobralSantos2018a}. Despite the progress, our picture is still  incomplete and   ongoing  and  future  surveys, such as \textit{The Extended Baryon Oscillation Spectroscopic Survey} \citep[eBOSS,][]{Dawson2016}, \textit{Dark Energy Spectroscopic Instrument} \citep[DESI,][]{DESI2016}, \textit{Euclid} \citep{Laureijs2011}, \textit{Wide Field Infrared Survey Telescope} \citep[WFIRST,][]{Dressler2012}, and   \textit{eROSITA} \citep{Merloni2012}, could soon fill the gaps. To optimally exploit the data from these upcoming galaxy surveys, synthetic galaxy catalogues are needed  \citep[e.g.][]{Blaizot2005, Kitzbichler2007, Guo2011, Merson2013, Lacey2016, Merson2018}. By using these mock catalogues it is possible to estimate uncertainties in deriving a given galaxy property, study selection effects, or quantify the impact of different sources of errors. In addition, it is possible to modify various assumptions regarding galaxy formation physics, and explore their impact on observable galaxy properties. Thus, realistic and  physically motivated mock catalogues are extremely important to interpret observational data in terms of the underlying galaxy-formation physics.\\

In particular, mock galaxy catalogues are particularly important for interpreting surveys that combine broad-band with narrow-band photometry \citep{Wolf2003,Moles2008, Ilbert2009, Perez-GonzalezPG2013, Benitez2014, Cenarro2018, Padilla2019}. These surveys attempt to inherit the power of spectroscopy in reliably estimating physical properties of galaxies, and of photometry in measuring the light in a spatially resolved manner while avoiding the pre-selection of targets. Thus, they deliver smaller statistical uncertainties and weaker degeneracies in estimating physical properties of galaxies compared to broad-band surveys. On the other hand, due to the complexity of the data and its acquisition, they might contain more uncertainties related to the measurement of line emission compared to spectroscopic surveys.\\

There are several requirements for realistic mock catalogues. First,   a galaxy formation model is needed that predicts all the relevant observable properties of galaxies, such as position, redshift, metallicity, stellar mass, or star formation rate \citep{Croton2006, Somerville2008, Guo2011, Lacey2016, Henriques2015}. Second, it is important to include emission lines from star-forming regions and quasars \citep{Orsi2014, Molino2014,ChavesMontero2017,Comparat2019}; although lines contribute in a relatively minor way to broad-band magnitudes, they can dominate the total flux in narrow and medium bands \citep[see e.g.][]{Sobral2009, Sobral2013, VilellaRojo2015, Matthee2015, Stroe2015, Stroe2017,SobralSantos2018b}. Third, it is necessary to project the light and spatial distribution of mock galaxies onto the observer's frame of reference. This, the so-called lightcone, is a crucial ingredient since a given narrow band can receive contributions from multiple emission lines at different redshifts.\\


During recent years various galaxy lightcones using merger trees of dark matter $N$-body simulations and galaxy formation models have been developed \citep[][]{Blaizot2005,KitzbichlerWhite2007,Merson2013,Overzier2013}. These mocks lightcones were designed for broad-band surveys, such as SDSS, where the contribution of emission lines in the final galaxy photometry was neglected. With the advancement of more sophisticated narrow-band photometric surveys such as \textit{Survey for High-z Absorption Red and Dead Sources} \citep[SHARDS,][]{Perez-GonzalezPG2013}, \textit{Javalambre-Photometric Local Universe Survey}\footnote{www.j-plus.es} \citep[J-PLUS,][]{Cenarro2018}, \textit{Javalambre Physics of the Accelerating Universe Astrophysical Survey} \citep[J-PAS,][]{Benitez2014}, and  \textit{Physics of the Accelerating Universe} \citep[PAU,][]{Padilla2019} the line contributions from star-forming galaxies need to be taken into account.  To date, few works have addressed this. For instance, \cite{Merson2018} by using the \texttt{CLOUDY} photo-ionisation code \citep{Ferland2013} included the {\Ha} emission in the \texttt{GALACTICUS} galaxy formation model \citep{Benson2012}. By constructing a 4 square degree catalogue they were able to predict the expected number of {\Ha} emitters as a function of redshift, a critical aspect for \textit{Euclid} and \textit{WFIRST} surveys. On the other hand, \cite{Stothert2018} performed forecasts for PAU employing the \texttt{GALFORM} version of \cite{Gonzalez-Perez2014} where the modelling of {\Ha}, {\oii}, and {\oiii} lines was included.\\


In this paper we present a new procedure to generate  synthetic galaxy lightcones, specially designed for narrow-band surveys. We employ state-of-the-art theoretical galaxy formation models applied to a large $N$-body simulation to predict the properties and clustering of galaxies. We improve these results with a model for the nebular emission from star-forming regions considering the contribution of nine different transition lines. The properties of these lines are computed separately for each mock galaxy based on its predicted star formation and metallicity. This is one of the first times that multiple emission lines have been included in mock galaxy lightcones following a self-consistent physical model \citep{Merson2018,Stothert2018}. Additionally, we embed the lightcone building procedure inside the galaxy formation modelling, allowing us to minimise the time-discreteness effects. As an application of our lightcone construction, we   generated catalogues for the photometry of the ongoing J-PLUS photometric survey \citep{Cenarro2018}   by observing thousands of square degrees of the northern sky with a specially designed camera of 2 $\rm deg^2$ field of view (0.55" $\rm pix^{-1}$ scale) and the unique combination of five broad-band (\usdss,\gsdss,\rsdss,\isdss,\zsdss) and seven medium- and narrow-band filters \citep[see Table 3 of][]{Cenarro2018}. We employed our mocks to test the capabilities of the survey in identify,  with a simple three-filter method (3FM), a population of emission-line galaxies at various redshifts. Specifically, all the emission lines that fall in a narrow-band filter centred at the $\rm H_{\alpha}$ rest wavelength ({\jha} filter). We showed how the $4000\AA$ break in the galaxy spectral energy distribution can cause an apparent excess, misidentified as line emission. However, we demonstrated that all significant excess (larger than 0.4 magnitudes) can be unambiguously attributed to emission lines\footnote{The mock catalogue is publicly available at  \href{https://www.j-plus.es/ancillarydata/mock\_galaxy\_lightcone}{https://www.j-plus.es/ancillarydata/mock\_galaxy\_lightcone}}.\\

This paper is organised as follows. In section \ref{sec:methodology} we describe the methodology we follow to construct the galaxy lightcone, predict galaxy properties, and model the strength of emission lines. In section \ref{sec:validation} we present various comparisons with observations, which illustrate the accuracy of our predictions. In section \ref{sec:apps} we employ our synthetic catalogues to study the selection of emission-line galaxies (ELGs) in J-PLUS. Finally, in section \ref{sec:conclusions} we summarise our main findings. In this work magnitudes are given in the AB system. A lambda cold dark matter $(\Lambda$CDM) cosmology with parameters $\Omega_{\rm m} \,{=}\, 0.25$, $\Omega_{\rm \Lambda}  \,{=}\, 0.75$, and $H_0 \rm \,{=}\, 73 \,km\;s^{-1}\,Mpc^{-1}$ is adopted throughout the paper. 
\section{Methodology}
\label{sec:methodology}

In this section we discuss the general procedure used to construct our mock galaxy lightcone. We start by describing our $N$-body simulation, galaxy formation models, and prescription for emission lines. Then we discuss our method for projecting simulated galaxy properties onto the lightcone.

\subsection{The Millennium simulation}
\label{subsec:MS}

The backbone of our lightcone construction is the \texttt{Millennium} $N$-body simulation \citep{Springel2005} which follows the cosmological evolution of $\rm 2160^3 \,{\backsimeq}\,10^{10}$ dark matter (DM) particles of mass $8.6 \,{\times}\,10^8\, \Msun$ inside a periodic box of 500 ${\rm Mpc}/h$ on a side, from $z\,{=}\,127$ to the present. The cosmological parameters used in the simulation were: $\Omega_{\rm m} \,{=}\, 0.25$, $\Omega_{\rm \Lambda}  \,{=}\, 0.75$, $\sigma_{8}\,{=}\,0.9$, $H_0 \rm \,{=}\, 73 \,km\;s^{-1}\,Mpc^{-1}$, $n\,{=}\,1$. Simulation data were stored at 63 different epochs (referred to as \textit{snapshots}) spaced logarithmically in time at early times ($z\,{>}\,0.7$) and linearly in time afterwards ($\rm \Delta t \,{\sim}\, 300 \,Myr$). At each snapshot, DM haloes and subhaloes were identified with a friends-of-friends (FoF) group-finder and an extended version of the $\rm SUBFIND$ algorithm \citep{Springel2001}. Objects more massive than $\mathrm{M_{halo}}\,{=}\,2.7\,{\times}\,10^{10}\, \Msun$ (corresponding to 32 particles) were kept in the catalogues. Subhaloes were linked across snapshots by tracking a fraction of their most bound particles, weighted by particle rank in a list sorted by binding energy. Subhalo catalogues and descendant links were arranged to form \textit{merger trees}, which allowed us to follow the assembly history of any given DM object. Given the halo mass resolution of the \texttt{Millennium} simulation, we expect converged properties and abundance for galaxies with stellar masses above $\mathrm{M_{stellar}} \,{\sim}\,10^{8}\, \Msun$ \citep[see][]{Guo2011,Henriques2015}.

\subsection{Galaxy formation model}
\label{subsec:SAM}
We employ a semi-analytical model (SAM) of galaxy formation to predict the properties of galaxies in our simulation. The aim of a SAM is to simulate the evolution of the galaxy population as a whole in a self-consistent and physically motivated manner. For this, galaxy properties such as star formation rate, stellar mass, luminosity, and magnitudes are a result of a simultaneous modelling of multiple physical processes, which typically include gas cooling, star formation, AGN and supernova feedback, metal enrichment, black hole growth, and galaxy mergers \citep[see e.g.][]{Bower2006,Guo2011,Gargiulo2015,Lacey2016}. All these processes are implemented through a system of coupled differential equations solved along the mass assembly history of DM objects, given by their respective merger tree \citep[see][for a review]{Baugh2006}.\\

In this work, we employ the \texttt{L-Galaxies} SAM code, in the well-tested variant presented by \cite{Guo2011}. In the future, we plan to apply the same procedure on the most updated version of \texttt{L-Galaxies}, presented in \cite{Henriques2015}. 
For completeness, below we summarise the main ideas and physical processes implemented in the model. A more thorough description of the model can be found in \cite{Croton2006,DeLuciaBlaizot2007,Guo2011,Henriques2015}.\\
Following the standard \cite{WhiteFrenk1991} approach, the \texttt{L-Galaxies} model assumes that when a DM halo collapses, a cosmic abundance of baryons collapses with it in the form of diffuse pristine gas, forming a quasi-static \textit{hot gas atmosphere}. Gradually, this gas cools and reaches the halo centre via cooling flows. As soon as the gas is accreted and cooled, the galaxy develops a cold gas disc which eventually triggers a \textit{secular burst} of star formation \cite{Guo2011}. Shortly after any star formation events, a fraction of new stars explode as a supernovae, enriching the environment with newly formed heavy elements and releasing an amount of energy able to eject and warm up the cold gas from the galaxy disc. The stellar feedback is not the only mechanism used to regulate the growth of the cold gas disc. For massive systems, the model uses the feedback from the central black hole (BH) to decrease the cooling rate, and hence stops the galaxy growth \citep{Croton2006}.
\\ 

Regarding the global galaxies properties, mergers, and secular evolution play an important role in the model triggering star formation and bulge or disc growth. On the merger side, the model distinguishes between two types of galaxy interactions. When the total baryonic mass of the less massive galaxy exceeds a fraction of the more massive one, a \textit{major merger} takes place. Otherwise it is a \textit{minor merger}. After a major interaction the discs of both galaxies are completely destroyed and the remnant galaxy is a pure spheroidal; instead, in a minor merger the remnant retains the stellar disc of the large progenitor and its bulge gains only the stars from the smaller progenitor. In both merger types the descendant galaxy undergoes a star formation process, known as \textit{collisional starburst} \citep{Somerville2001}, whose feedback process is the same as the secular star formation. For the galaxy secular evolution, the code takes into account the disc instabilities (DIs). In this context, DIs refers to the process by which the stellar disc becomes massive enough to be prone to non-axisymmetric instabilities, which ultimately lead to the formation of a central ellipsoidal component via the buckling of nuclear stellar orbits. The criterion used for modelling the disc instabilities is an analytic stability test based on \cite{Mo1998}. When the instability criterion is met, the code transfers the sufficient stellar mass from the disc to the bulge to make the disc marginally stable again \citep[see][]{IzquierdoVillalba2019}.\\

With respect to the dust modelling, the \texttt{L-Galaxies} model follows the \cite{DeLuciaBlaizot2007} formalism, which  considers separately the extinction coming from the diffuse interstellar medium and that from the molecular birth clouds within which stars are formed.\\

Finally, regarding the \textit{large-scale} effects, the model includes processes such as \textit{ram pressure} or \textit{tidal interactions} that can completely remove the hot gas atmosphere around satellite galaxies and eventually destroy their stellar and cold gas components \citep{Guo2011,Henriques2015}.\\

\begin{figure}
        \centering
        \includegraphics[width=1.\columnwidth]{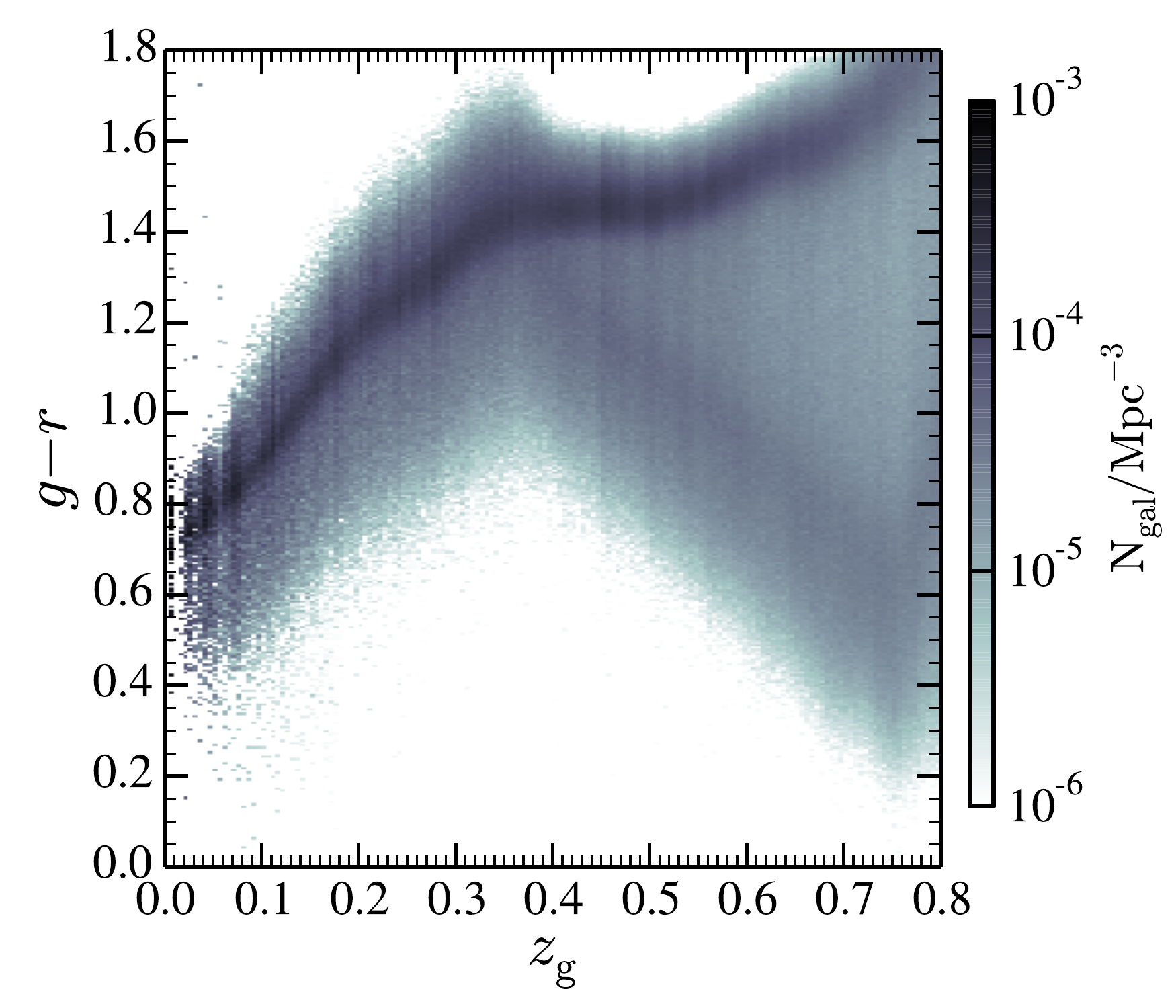}
        \caption{For our mock galaxies with $\rm M_{stellar}\,{>}\,10^{10} M_{\odot}/\mathit{h}$, the observed colour {\gsdss-\rsdss} as a function of redshift ($z_{\rm g}$). The grey scale represents the number density of galaxies (darker regions corresponds to larger number densities).}
        \label{fig:color_z}
\end{figure}



\subsection{Lightcone construction} \label{subec:Ligthcone_construction}


We now outline our method for constructing a lightcone. We start by defining the location of an observer and specifying the orientation, geometry, and angular extent of the lightcone. Then we define how we identify the moment when a galaxy crosses the observer past lightcone.\\

\begin{figure*}
        \centering
        \includegraphics[width=1.\columnwidth]{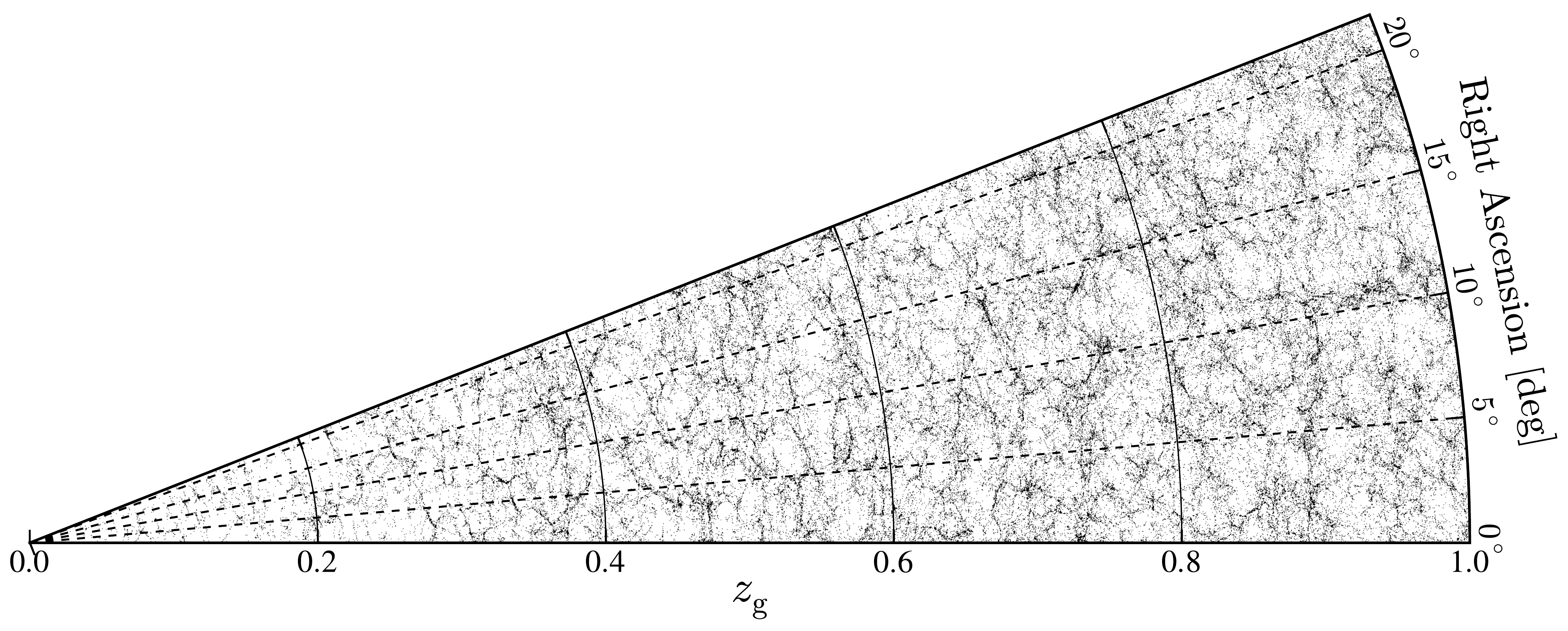}
        \includegraphics[width=1.\columnwidth]{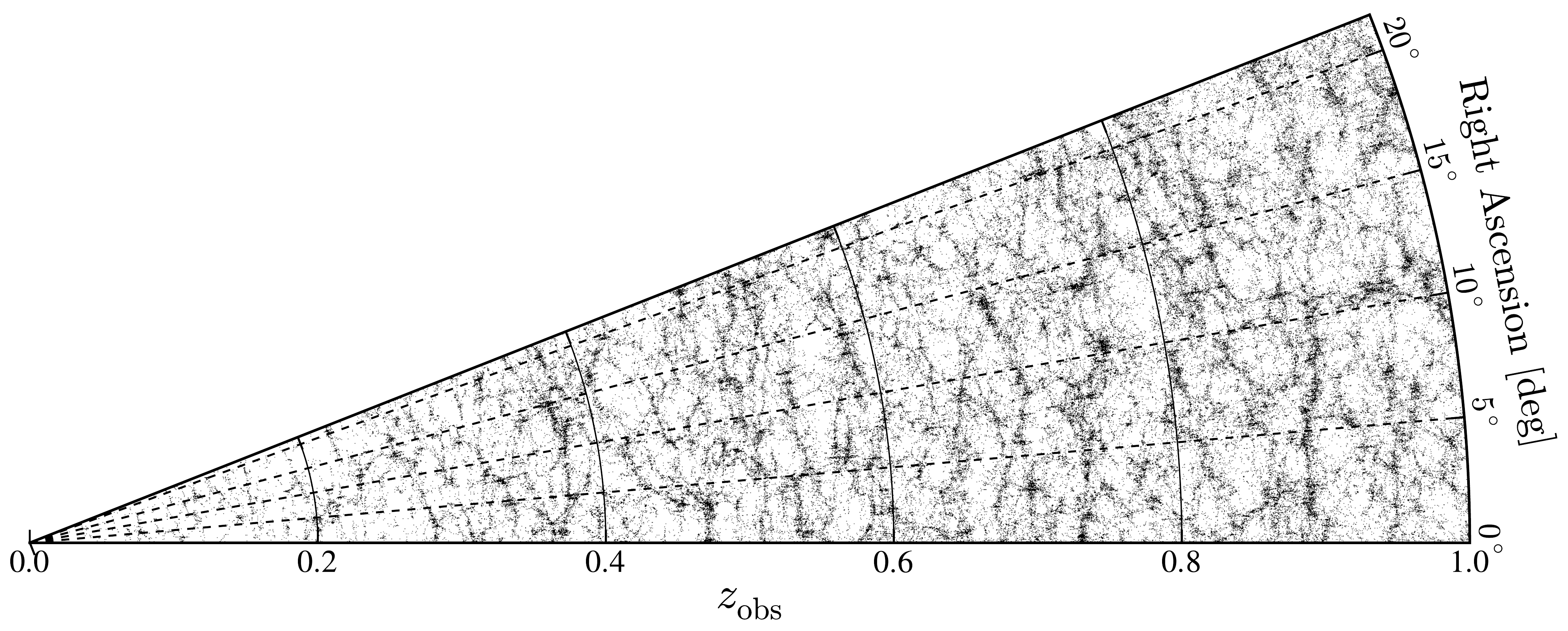}
        \caption{\textit{Left panel}: Spatial distribution of galaxies in a thin angular slice ($1\deg$) inside our mock lightcone. Each galaxy with $\mathrm{M_{stellar}}\,{>}\,10^{10}\,\Msun$ is shown as a black
                dot. The contribution of peculiar velocities is not included. \textit{Right panel}: The same, but including the peculiar velocities in the estimation of the redshift. The clustering   is enhanced on large scales due to coherent bulk motions, whereas it is damped on small scales due to random motions inside dark matter haloes.}
        \label{fig:LightCone}
\end{figure*}
The 500 Mpc/$h$ side-length of the \texttt{Millennium} simulation is not always able to encompass the full volume, or redshift range, of observational surveys. Thus, to cover the relevant regions we take advantage of its periodic boundary conditions and replicate the simulated box eight times in each coordinate direction. This corresponds to a maximum redshift of $z\,{\sim}\,3$, and will also allow us to incorporate high-$z$ ELGs as potential contamination for low-$z$ ELGs.\footnote{We apply our lightcone construction in the \texttt{Millennium} merger trees rather than in the \texttt{Millennium-XXL} because of the coarser mass resolution of the latter (with a particle mass of $\rm {\sim}\,10^9 M_{\odot}/\mathit{h}$). The minimum halo mass of \texttt{Millennium-XXL} would cause completeness effects in the magnitude range explored in this work and would impose a  line luminosity threshold that is too high for us to trust our results (e.g. $\rm {\gtrsim}\,10^{41}\,erg/s$ at $z\,{\sim}\,0$ for \Ha line, see \citealp{Orsi2014}).} Although the replicated volume underestimates the total number of independent Fourier modes in a survey, it is adequate when different redshift slices are considered separately and the redshift direction is chosen appropriately, as we discuss below.\\

For convenience, we place the observer at the origin of the first replication, and define the extent of the lightcone as the angular size of $1000\,{\rm Mpc}/h$ at $z\,{\sim}\,1$. This ensures that no more than two repetitions are required to represent the cosmic structure in any redshift shell up to $z\,{\sim}\,1$. This provides a $22.5\degree \,{\times}\,22.5\,{\degree}$ (${\sim}\,309.4\deg^2$) lightcone. The orientation of the lightcone was chosen following \cite{KitzbichlerWhite2007} to minimise repetition of structure along the line of sight (LOS). According to their methodology, the LOS  passes by the first periodic image at the point $\rm (\mathit{n}L,\mathit{m}L, \mathit{nm}L)$, where $m$ and $n$ are integers with no common factor and $\rm L $ is the box size. We set the values of $n\,{=}\,2$ and $m\,{=}\,3$, resulting in a viewing direction $\rm (\theta,\varphi)\,{=}\,(58.9\degree,56.3\degree$). In Appendix~\ref{appendix:repetition}, we show that this LOS yields a small overlap between box replicas.\\


The next step is to determine the moment when galaxies cross the observer's past lightcone. There are several different methods in the literature for this \citep[e.g.][]{KitzbichlerWhite2007,Merson2013}, most of which interpolate galaxy properties across the discrete dark matter simulation snapshots or by directly storing the DM mass field as the N-body simulation evolves. Here we have decided to follow a different approach. \texttt{L-Galaxies} accurately follows in time the evolution of individual galaxies between DM snapshots with a time step resolution $\rm {\lesssim}\, 5\,{-}\,19 \, Myr$. This includes the tracking of central, satellite, and orphan galaxies (i.e. those whose DM host has fallen below the resolution of the simulation), improving along the way the links of the underlying subhalo merger tree \citep[][]{DeLuciaBlaizot2007}. Here we take advantage of this and use the \textit{galaxy merger trees} as an estimation of the continuous path in space-time of a galaxy. Linearly interpolating between two contiguous galaxy time steps, we search for the the lightcone crossing redshift, $z_{\rm g}$, where the comoving radial distance is equal to the distance to the observer. The galaxy properties are then evolved down to that exact moment inside \texttt{L-Galaxies}. This approach has the advantage of reducing an artificial discretisation of galaxy properties usually seen in lightcone algorithms \citep[see e.g. Fig. 4 in][]{Merson2013}. To illustrate that discretizaton effects in our mock galaxy photometry are small, in Fig.~\ref{fig:color_z} we show the observed colour {\gsdss-\rsdss} as a function of redshift \citep[which is usually the most affected quantity, see][]{Merson2013} for galaxies with $\rm M_{stellar}\,{>}\,10^{10} M_{\odot}/\mathit{h}$. No evident discontinuities are seen along the {\gsdss-\rsdss} axis.\\

Finally, we add the contribution of peculiar velocities to the observed redshift of a galaxy as
\begin{equation}
\label{equation:z_obs}
z_{\rm obs} = (1+z_{\rm g}) \left(1 + \frac{v_{r}}{c} \right) - 1  \; ,\end{equation}
\noindent where $z_{\rm g}$ is the geometrical redshift at which the galaxy crosses the lightcone and $v_r$ is the LOS component of its peculiar velocity, and $c$ is the speed of light.\\

The spatial distribution of galaxies in our lightcone inside a  ${1\,\deg}$ slice  is presented in  Fig.~\ref{fig:LightCone}. We only display galaxies more massive than $10^{10}\,\Msun$. No visible discreteness effects originating from a finite number of simulation snapshots are seen. 

\subsection{Line emission modelling}
\label{subsec:LineModelling}

In order to include the contribution of emission lines to the predicted photometry of our mock galaxies, we follow the model described in \cite{Orsi2014}. Specifically, we consider the contribution of nine different lines: $\rm Ly{\alpha}$ (1216\AA), \Hb (4861\AA), \Ha (6563\AA), {\oii} (3727\AA, 3729\AA), {\oiii} (4959\AA, 5007\AA),  $\rm [\ion{Ne}{III}]$ (3870\AA), {\oi} (6300\AA), $\rm [\ion{N}{II}]$ (6548\AA, 6583\AA), and $\rm [\ion{S}{II}]$ (6717\AA, 6731\AA), 
which are those we expect to contribute most significantly to the rest-frame optical wavelength.\\

In brief, the \cite{Orsi2014} model obtains the lines flux based on a \cite{LevesqueKewleyLarson2010}\footnote{These authors computed the theoretical  SEDs for $\rm \ion{H}{II}$ regions using the \texttt{Starburst99} code \citep{Leitherer1995} in combination  with the \texttt{MAPPINGS-III} photo-ionisation code \citep{Dopita1995,Dopita1996,Groves2004}.} model grid of $\rm \ion{H}{II}$ region. Four different parameters are needed as an input to the grid: (i) age of the stellar cluster that provides the ionising radiation ($t_*$), (ii) density of the ionised gas ($\mathrm{n}_{e}$), (iii) galaxy gas-phase metallicity ($\rm Z_{cold}$), and (iv) ionisation parameter ($\rm q$). For the first two parameters we assume constant values:  $t_{*}\,{=}\,0$ and $\mathrm {n}_{e} \mathrm{{=}\, 10 \, cm^{-3}}$ \citep[see the discussion in][]{Orsi2014}. The last two parameters are directly set by the cold gas metallicity predicted by our galaxy formation model adopting the following relation for the ionisation parameter,\begin{equation}\label{equation:RelationZq}\rm  q\left(Z\right) = q_0 \left(\frac{Z_{cold}}{Z_0}\right)^{-\gamma} \,\, [cm/s],
\end{equation}
\noindent where $\rm q_0$, $\rm Z_0$, and $\gamma$ are free parameters set to $\rm 2.8\,{\times}\,10^7\, cm/s$, 0.012, and 1.3, respectively, to match observational measurements of {\Ha}, {\oii}, and {\oiii} luminosity functions \citep{Orsi2014}.\\

By using the predicted line fluxes, the luminosity of a given line, $\rm L(\lambda_j)$, is given by\begin{equation}\label{equation:luminosities}
\rm  L(\lambda) \,{=}\,  1.37\times 10^{-12} Q_{H^o} \frac{F(\lambda_j|\,q,Z_{cold})}{F(H_{\alpha}|\,q,Z_{cold})} \;\;\; [erg/s],
\end{equation}

\noindent where $\rm F(H_{\alpha}|\,q,Z_{cold})$ and $\rm F(\lambda_j|\,q,Z_{cold})$ are respectively the flux of \Ha and $\lambda_j$ line in a galaxy with ionisation parameter q and metallicity $\rm Z_{cold}$ and $\rm Q_{H^o}$ is the ionisation photon rate in units of $\rm s^{-1}$ calculated from the galaxy instantaneous star formation rate predicted by our SAM. Here we assume that all the emitted photons contribute to the production of emission lines.\\ 

We note that the model predictions for the Baldwin, Phillips \& Telervich diagram \citep[BPT diagram,][]{Baldwin1981}\footnote{The BPT diagram of the model can be found in \cite{Orsi2014} Figure 1, with $\gamma\,{=}\,1.3$.} and for the evolution of the emission-line luminosity function are in  reasonable agreement with the observations. For more information, we refer the reader to \cite{Orsi2014}.

\subsection{Observed magnitudes}\label{subsection:comp_magnitudes}
Once we  had placed galaxies in the lightcone and computed their physical properties (such as stellar mass, star formation rate, and metallicity), we derived their observed photometric properties. The observer-frame apparent magnitudes in the AB system, $m_{AB}$, are defined as 
\begin{equation} \label{eq:f_nu_cont}
m_{AB} {=} -2.5\log_{10} \left( f_{\rm \nu} \right) - 48.6  
         {=} -2.5 \log_{10} \frac{\int \mathrm{T(\lambda) \lambda} f_{\rm \lambda} \mathrm{d}\lambda}{\rm c \int \frac{T(\lambda)} {\lambda} d\lambda} - 48.6,         
\end{equation} 
\noindent where c is the speed of light, $\rm T(\lambda)$ is the filter transmission
curve,  $\rm \lambda$ is the wavelength,  $f_{\nu}$ is the flux density per frequency interval $\mathrm{ erg\,s^{-1}\,cm^{-2}\,Hz^{-1}}$, and $f_{\lambda}$ is the flux density per wavelength interval in $\mathrm{ erg\,s^{-1}\,cm^{-2}\,\AA^{-1}}$.\\

We assume that our final galaxy photometry is the sum of two contributions. The first  is the continuum emission from the mixture of stellar populations hosted by the galaxy ($f_{\lambda}^{c}$). The second  is the specific flux of all the recombination lines ($f_{\lambda}^{l_j}$) generated in $\rm {H_{\,II}}$ regions. Therefore, 
\begin{equation}\label{eq:total_flux}
f_{\lambda} \equiv f_{\lambda}^{c} + \mathrm{\sum_{\mathit{j}=1}^{n_{lines}}} f_{\lambda}^{l_j}.
\end{equation}
\noindent Including Eq.~\eqref{eq:total_flux} in Eq.~\eqref{eq:f_nu_cont}, the magnitude $m_{AB}$ can be expressed as


\begin{equation}
\label{equation:f_nu_cont_split}
m_{AB} = {2.5 \log_{10}}\left[ 10^{-0.4(m_{AB}^c +48.6)} + \frac{\int \mathrm{T(\lambda) \lambda} \sum_{j=1}^{\mathrm{n_{lines}}}  f_{\lambda}^{l_j} d\lambda}{\rm c \int \frac{T(\lambda)} {\lambda} d\lambda}\right]-48.6.
\end{equation}

The magnitude $m_{AB}^c$ is computed by our SAM in a self-consistent way according to the galaxy evolution pathway (Section~\ref{subsec:SAM}). By using the \cite{Bruzual2003} synthesis models and a Chabrier initial mass function, \texttt{L-Galaxies} updates the galaxy luminosity in each photometric band every time that the galaxy experiences a star-forming event. Additionally, a model is assumed to account for the light attenuation due to the absorption in the interstellar medium (ISM) and molecular clouds \citep[see][]{DeLuciaBlaizot2007,Guo2011,Henriques2015}. When the galaxy crosses the lightcone, the observed magnitudes in the chosen photometric system are output according to the total luminosity at that moment.\\

The contribution of emission lines (second term in Eq.~\ref{equation:f_nu_cont_split}) is taken into account in post processing. Throughout this work we only consider the line emission produced by star formation events, ignoring the contribution of AGNs. The line ${l_j}$ is added in its observed wavelength with $f_{\lambda}^{l_j}$ determined by a $\delta$-Dirac profile of amplitude $\rm F(\lambda_j|q,Z_{cold})$. In our case, $\rm n_{lines}$ are the nine lines described in Section~\ref{subsec:LineModelling}. As in the case of the galaxy continuum, all the line fluxes included here are affected by the surrounding dust. A discussion of the impact and modelling of dust is presented in Section \ref{subsec:linecounts}.


\begin{figure*}
        \centering
        \includegraphics[width=\textwidth]{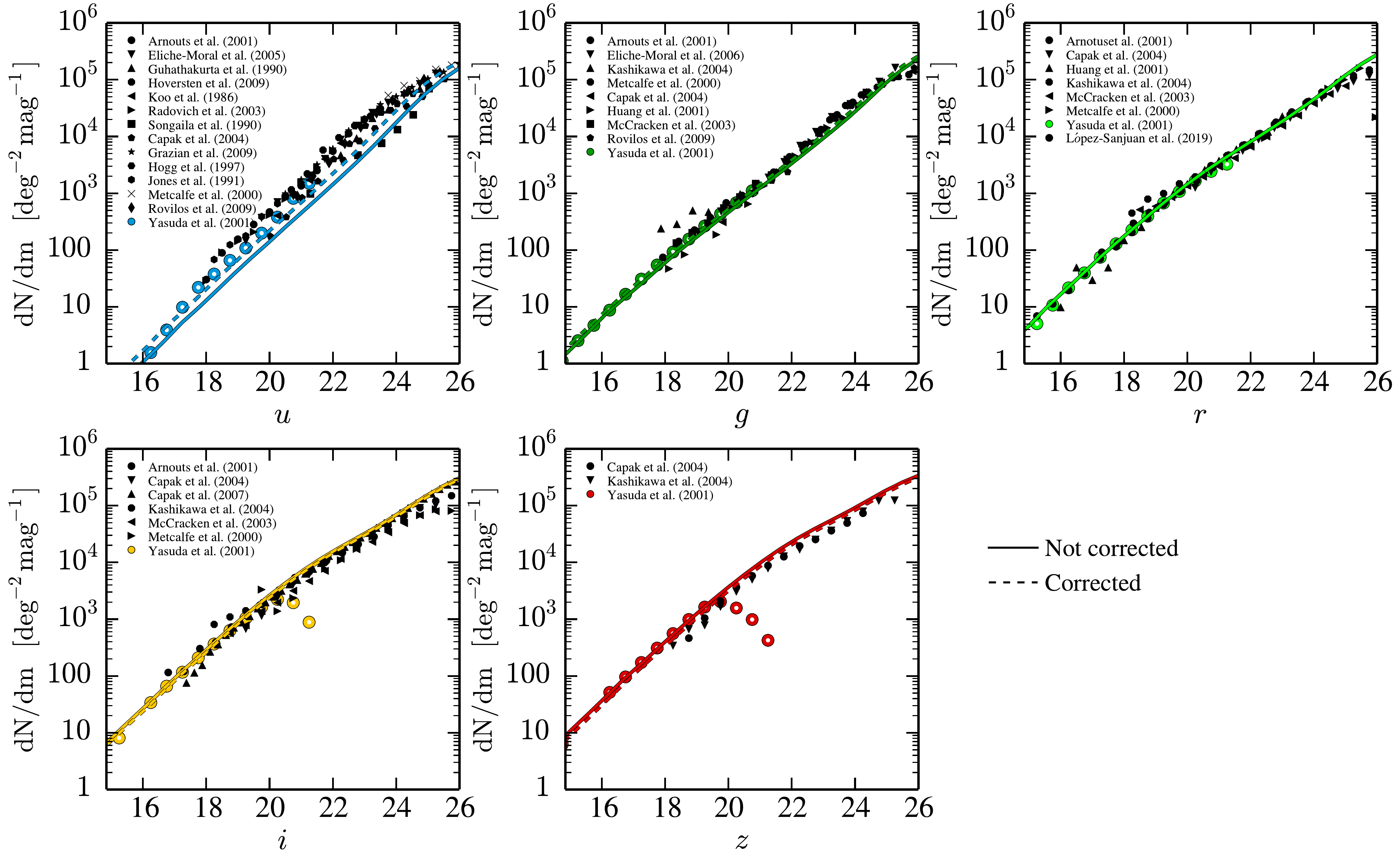}
        \caption{Abundance of galaxies as a function of observed magnitude for the different SDSS bands. Symbols show a compilation of observational results from the different datasets, with the measurements from SDSS highlighted as large coloured circles \protect{\citep{Yasuda2001}}. Shown in  each panel are the predictions from our mock galaxy lightcone before and after a correction to apparent magnitudes (solid and dashed lines, respectively) designed to improve the agreement with observations (see Section~\ref{subsec:counts}). The drop in the \protect{\cite{Yasuda2001}} data in the \isdss and \zsdss SDSS bands at faint magnitudes is caused by selection effects. The sample used in that work was selected in \rsdss with a magnitude range of $12\,{\leq}$\rsdss${\leq}\,21$. }
        \label{fig:NumberCounts}
\end{figure*}

\section{Validation} \label{sec:validation}

In this section we present a set of basic tests for our mock galaxy lightcone. In Section \ref{subsec:counts} we compare the predicted galaxy number counts against a compilation of observations in five broad bands ({\usdss}, {\gsdss}, {\rsdss}, {\isdss}, {\zsdss}). In Section  \ref{subsec:linecounts}, we extend our comparison to the luminosity functions of \Ha, \Hb, {\oii}, and {\oiiiFd} lines. We use this comparison to calibrate a dust obscuration model. Finally, in Section~\ref{subsec:Favlole_clustering} we show the ability of our mock to reproduce the observed clustering of \gsdss-band selected galaxies.

\subsection{Galaxy number counts} \label{subsec:counts}

In Fig.~\ref{fig:NumberCounts} we show the total number of galaxies in our lightcone mock as a function of apparent magnitude. We present our predictions for the five SDSS broad-band magnitudes, and compare them to various observational estimates as indicated by the legend \citep{Koo1986, Guhathakurta1990, Jones1991, Hogg1997, Arnouts2001, Yasuda2001, Metcalfe2001, Huang2001, McCracken2003, Radovich2004, Kashikawa2004, Capak2004, Eliche-Moral2005, Capak2007, Hoversten2009, Rovilos2009, Grazian2009, LopezSanjuan2019}. We note that different observations usually employ slightly different filter transmission curves; however, this introduces only a very minor correction, which we ignore.\\ 

Our theoretical predictions and the observations are in good agreement, especially for the \rsdss, \isdss, and \zsdss\, SDSS bands. This represents an important validation of our methodology. Even so, we find a slight systematic disagreement across bands, transiting from  well-matched number counts on long wavelengths to an underestimation at short ones (\usdss and \gsdss). At such wavelengths the magnitudes are sensitive to the  rather crude dust modelling implemented in the SAM. Since our main goal is to create mock galaxy catalogues that are as realistic as possible, we have applied an ad hoc correction to our apparent magnitudes, 

\begin{equation} \label{eq:correction}
m_{AB} \rightarrow m_{AB} + \alpha \left( \frac{\lambda_{0}}{\lambda_{AB}} -1 \right),
\end{equation}

\noindent where $\lambda_{AB}$ is the effective wavelength of the filter under consideration, and $\alpha$ and $\lambda_0$ are free parameters that we set respectively to $-0.47$ and $6254\,{\AA}$ (the central wavelength of the {\rsdss} SDSS filter) by requiring an improved agreement with the number counts shown in Fig.~\ref{fig:NumberCounts}. We apply Eq.~\eqref{eq:correction} to all the photometric bands we consider, including narrow and intermediate bands not used in the calibration. Our updated predictions, displayed as dashed lines in Fig.~\ref{fig:NumberCounts}, are in  better agreement with the data for blue bands, which increases the overall level of realism of our lightcone.

\begin{figure*}
        \centering
        \includegraphics[width=1.6\columnwidth]{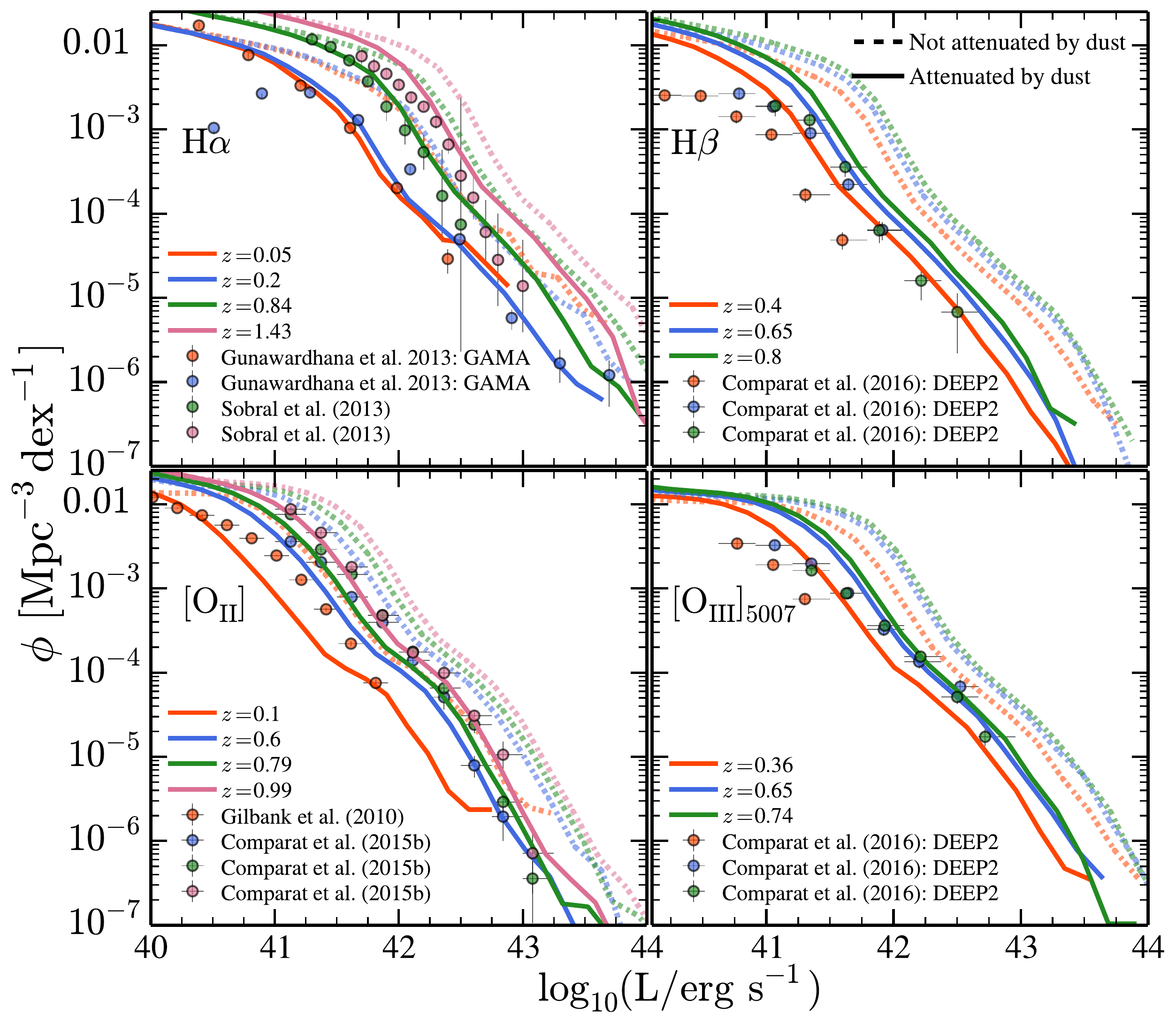}
        \caption{Luminosity function of emission lines. Each panel displays the results for a different emission line (from top to bottom and from left to right): {\Ha}, {\oii}, {\Hb}, and {\oiiiFd}. Shown in  each panel are the various observational estimates (cf.  Section~\ref{subsec:linecounts} and Appendix~\ref{appendix:LF}) at different redshifts, colour-coded as shown in each  legend. Dashed and solid lines display the results from our mock galaxy catalogue before and after applying an empirical model for dust attenuated, respectively. This dust model has been calibrated to improve the agreement between the observations and our predictions.}
        \label{fig:ELG_LF}
\end{figure*}

\subsection{Emission-line luminosity functions and line dust attenuation} \label{subsec:linecounts}

A distinctive feature of our mock lightcone is the inclusion of emission lines. Our model estimates line luminosities based on the \textit{intrinsic} amount of photons produced during an event of instantaneous star formation. However, star-forming galaxies are expected to also contain a large amount of dust, which can significantly attenuate the luminosity of these emission lines.  In the following we detail our dust-attenuation model, which is calibrated by making use of the well-constrained \Ha, \Hb, {\oii}, and {\oiiiFd} luminosity functions (LFs) provided by previous works\footnote{The comparison between observed and predicted LFs includes the small corrections due to the diverse cosmologies assumed by the different works. We checked that the variations in the LF amplitude due to this effect are minimum (${<}\,2$\%).}. We note that given the much more complex  physics involved in $\rm Ly{\alpha}$-photon radiative transfer in star-forming galaxies \citep{Gurung2019a,Gurung2019b,Weinberger2019}, we do not use $\rm Ly{\alpha}$ line luminosities functions for our calibrations.\\

In Fig.~\ref{fig:ELG_LF}, we present a comparison of our LF predictions against different observed (not dust corrected) luminosity functions for \Ha, \Hb, {\oii}, and {\oiiiFd} \citep{Gilbank2010,Gunawardhana2013,Sobral2013,Comparat2016}. 
For clarity, we only show four different redshifts and defer the comparison with other redshifts to  \hyperref[appendix:LF]{Appendix~\ref{appendix:LF}}. Our \textit{dust-free} predictions are, at all redshifts, above the observed ones. This suggest dust attenuation is required to match the observations. \\



\begin{figure} 
        \centering
        \includegraphics[width=\columnwidth]{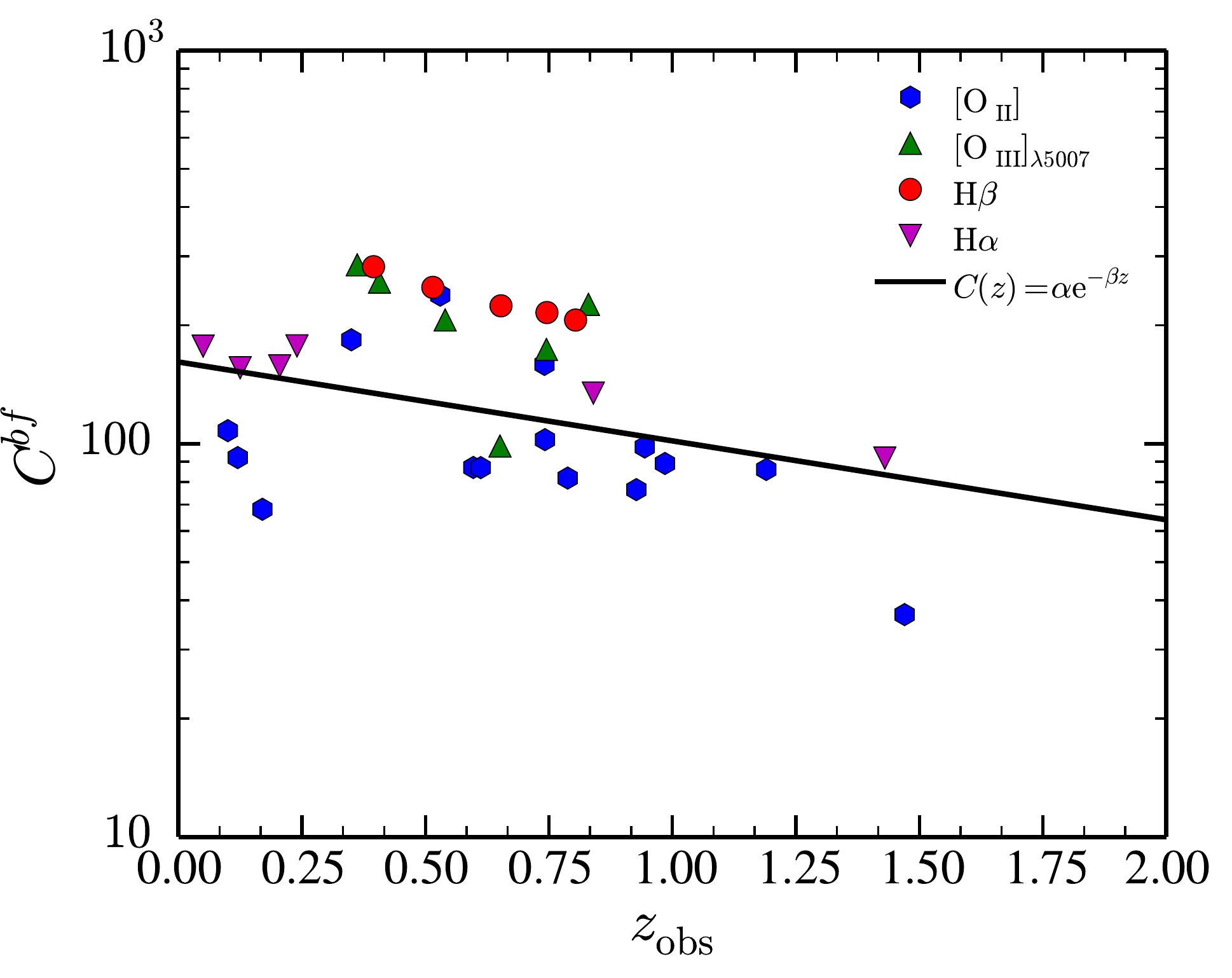}
        \caption{Redshift dependence of the dust attenuation coefficient, $C^{bf}$, applied to the
        nebular emission in our mock galaxies. Symbols represent the mean value estimated by 
        requiring agreement between the predicted and observed luminosity function for either 
        {\Ha}, {\Hb}, {\oii}, or {\oiiiFd} (violet, blue, green, and red, respectively). The best fit 
        relation is shown by the solid black line.} 
    \label{fig:Dust_correction}
\end{figure}

Given the difficulty in properly simulating dust formation and destruction \citep[e.g.][]{Fontanot2011}, here we resort to a simple empirical dust modelling. The goal is to consistently reproduce the observed luminosity functions across redshifts for different lines. Following \cite{DeLucia2004}, for each galaxy we compute a mean absorption coefficient as\begin{equation}
\label{equation:slabgeometry}
A_{\rm \lambda} =-2.5 \log_{10}\left(  \frac{1- \mathrm{e}^{-\tau_{\rm \lambda} \mathrm{sec}\,\theta}}{\tau_{\rm \lambda} \mathrm{sec}\,\theta}\right)  ,
\end{equation} 
where $\theta$ is the inclination angle of the galaxy with respect to the LOS (randomly chosen) and $\tau_{\rm \lambda}$ is the optical depth associated with stellar birth clouds. Here, we assume that $\tau_{\rm \lambda}$ has the following dependence on the cold gas metallicity of the host galaxy,\begin{equation} \label{equation:extinction}
\tau_{\rm \lambda} = C(z) \; \mathrm{Z_{cold}} \frac{\rm A_{V}}{\rm A_{B}}\frac{\rm A(\rm \lambda)}{\rm A_V} \; ,
\end{equation} 
where the values of $\rm A_{B}/A_{V}$ and $\rm A(\rm \lambda)/A_{V}$ are computed based on the  extinction curves of \cite{Cardelli1989} and $C(z)$ is a free  parameter, which we refer to as the \textit{dust attenuation coefficient}, that controls the amplitude and redshift dependence of our dust attenuation model. 

To constrain the value of $C(z)$, we compute the dust-attenuated luminosity function for a wide set of values for $C$ (namely $[0,1000]$). We then find the value of $C$ that minimises the root mean squared differences with the observed luminosity function. We apply this procedure separately for each line and redshift. The best-fit values for $C$ ($C^{bf}$) as a function of redshift for the {\Ha}, {\oii}, {\oiiiFd}, and {\Hb} lines is shown in Fig.~\ref{fig:Dust_correction},  which shows that the corrections for {\Hb} and {\oiiiFd} are systematically larger than those for { \Ha} and {\oii}. We connect this evidence to the slight overestimation of the galaxy intrinsic \Hb and {\oiiiFd} luminosity \citep[see][]{Orsi2014}. Because the refinement of \Hb and {\oiiiFd} modelling is beyond the scope of our work, we absorb these overestimations in our dust correction at the price of using large C coefficients for these lines.\\

We find that all three of the lines considered and the whole redshift range display a consistent behaviour, with a smaller attenuation coefficient required at high redshift. Our results are well described by $ C(z) \; \mathrm{= \alpha} e ^{\mathrm{-\beta} z}$, with $\rm \alpha\,{=}\,161.46 \pm 30.3$ and $\beta\,{=}\,0.46 \pm 0.23$. This relation is shown as a solid black line in Fig.~\ref{fig:Dust_correction}. We  employ it to model dust attenuation for every line we include in our mock, with the exception of {\Ha}. The resulting \textit{dust-attenuated} line luminosity functions in our mock lightcone are displayed in Fig.~\ref{fig:ELG_LF}. As intended, we find a better agreement with the observational measurements, which supports the validity of our mocks  for analysing and predicting emission lines in the Universe. Nevertheless, given the simplicity of our dust correction, the final results have their limitations. For instance, while the {\oii} line is slightly over-corrected at low-$z$, \Hb is under-corrected at high-$z$. In future work we will address a more sophisticated dust attenuation.

\subsection{Clustering of high redshift \gsdss-band selected sources} \label{subsec:Favlole_clustering}

\begin{figure}
        \centering
        \includegraphics[width=\columnwidth]{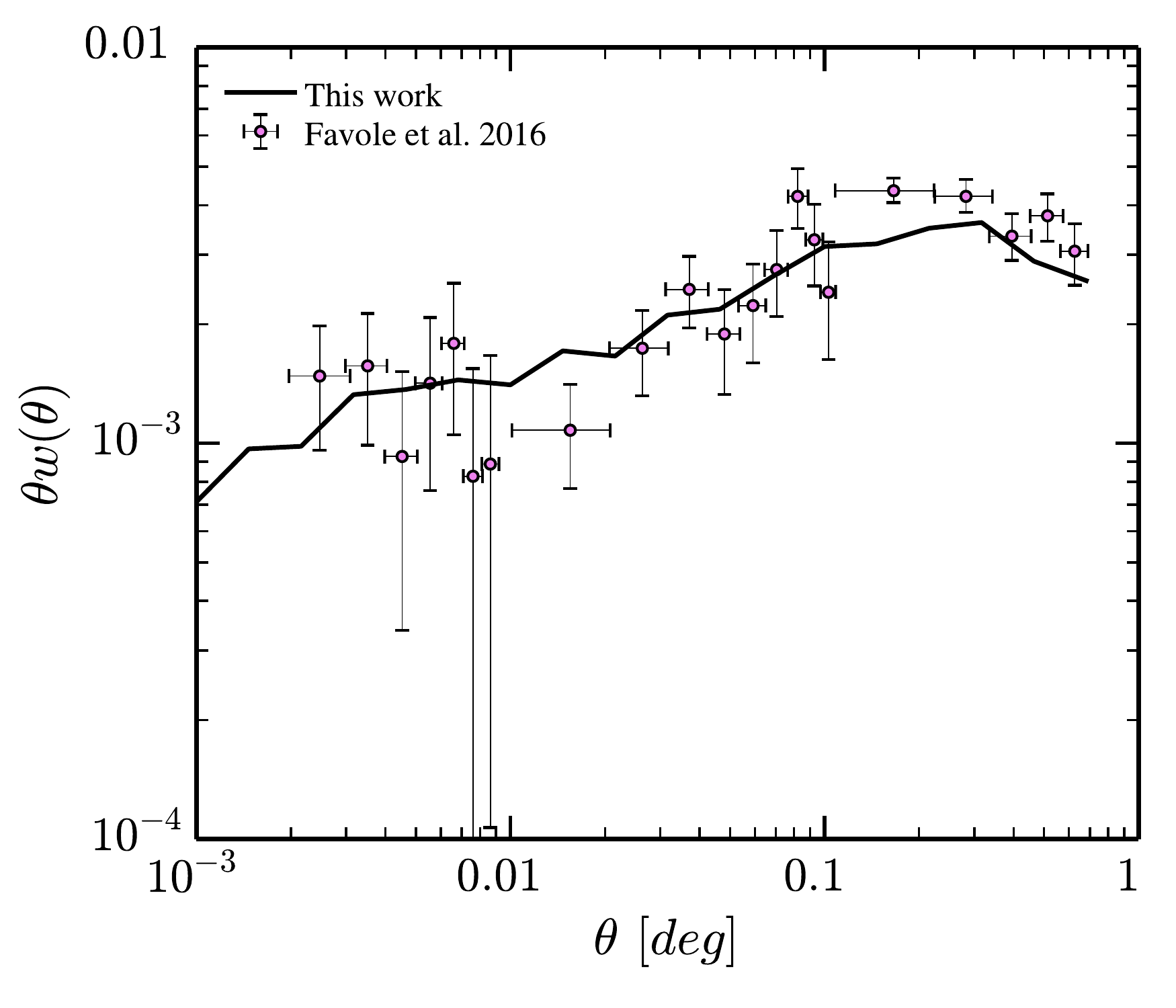}
        \caption{Two-point angular correlation function of \gsdss-selected galaxies (20 $<$ \gsdss $<$ 22.8) in the redshift range $0.6\,{<}\,z\,{<}\,1.0$. Purple symbols displays the measurements of \protect{\cite{Favole2016}} whereas solid black line shows the predictions of this work. We display $\theta \times w(\theta)$ to enhance the dynamic range shown. The clustering has been rescaled from the WMAP cosmology to the PLANCK one following \protect{\cite{Springel2018}}.}
        \label{fig:clustering_Favole}
\end{figure}

An important validation of our mock lightcone is the spatial distribution of ELGs. In this subsection we compare our results to the measurements of \cite{Favole2016} who computed the two-point angular correlation function of \gsdss-band and redshift-selected galaxies in SDSS (a selection designed to be a proxy for \oii \space emitter selection).\\ 

We construct a mock galaxy sample by applying the same selection criteria as in \cite{Favole2016}. Specifically, we impose $20\,{<}\,g\,{<}\, 22.8$ and $0.6\,{<}\,z\,{<}\,1.0$. This results in a sample of ${\sim}\,2{\times}10^5$ galaxies, with a median star formation rate of $\rm 1.70\,M_{\odot}/yr$. We compute the angular correlation function, $w(\theta)$, using the \texttt{Corfunc} package \citep{manodeep_sinha_2016_55161} with the Landy-Szalay estimator \citep{LandySzalay1993},

\begin{equation}
w(\theta) = \frac{DD(\theta) - 2DR(\theta) + RR(\theta)}{RR(\theta)},
\end{equation}

\noindent where $DD(\theta)$ is the number of galaxy-galaxy pairs within separation $\theta$, $RR(\theta)$ is the expected number of such pairs in a random sample generated with the same selection function of our mock data, and $RD(\theta)$ the data-random pairs.\\

In Fig.~\ref{fig:clustering_Favole} we present the comparison. The clustering has been rescaled from the WMAP cosmology of the \texttt{Millennium} to the PLANCK cosmology following \cite{Springel2018}. Our predictions display a remarkable agreement with the observations, being statistically consistent within the measurement uncertainties. Additionally, there is a good agreement in the physical properties of the underlying sample: our mock sample has a median host halo mass of $\rm M_{vir}^{FOF} \,{=}\, 1.249{\times}10^{12} M_{\odot}$ with a  $27\%$ satellite fraction. These figures are to be compared with a typical host halo mass of $\rm (1.25 \,{\pm}\,0.45){\times}10^{12} M_{\odot}$ and a satellite fraction of $\sim22.5\%$, as estimated by \cite{Favole2016}. We note that this level of agreement compares favourably with respect to what it is found in other SAMs \citep[e.g.][]{GonzalezPerez2018}.

\section{J-PLUS mock galaxy catalogues} \label{sec:apps}

In this section, we  employ our procedure to build lightcones to mimic the J-PLUS survey. We explore its ability to characterise ELGs in the Universe.\\

\begin{figure}
        \centering
        \includegraphics[width=\columnwidth]{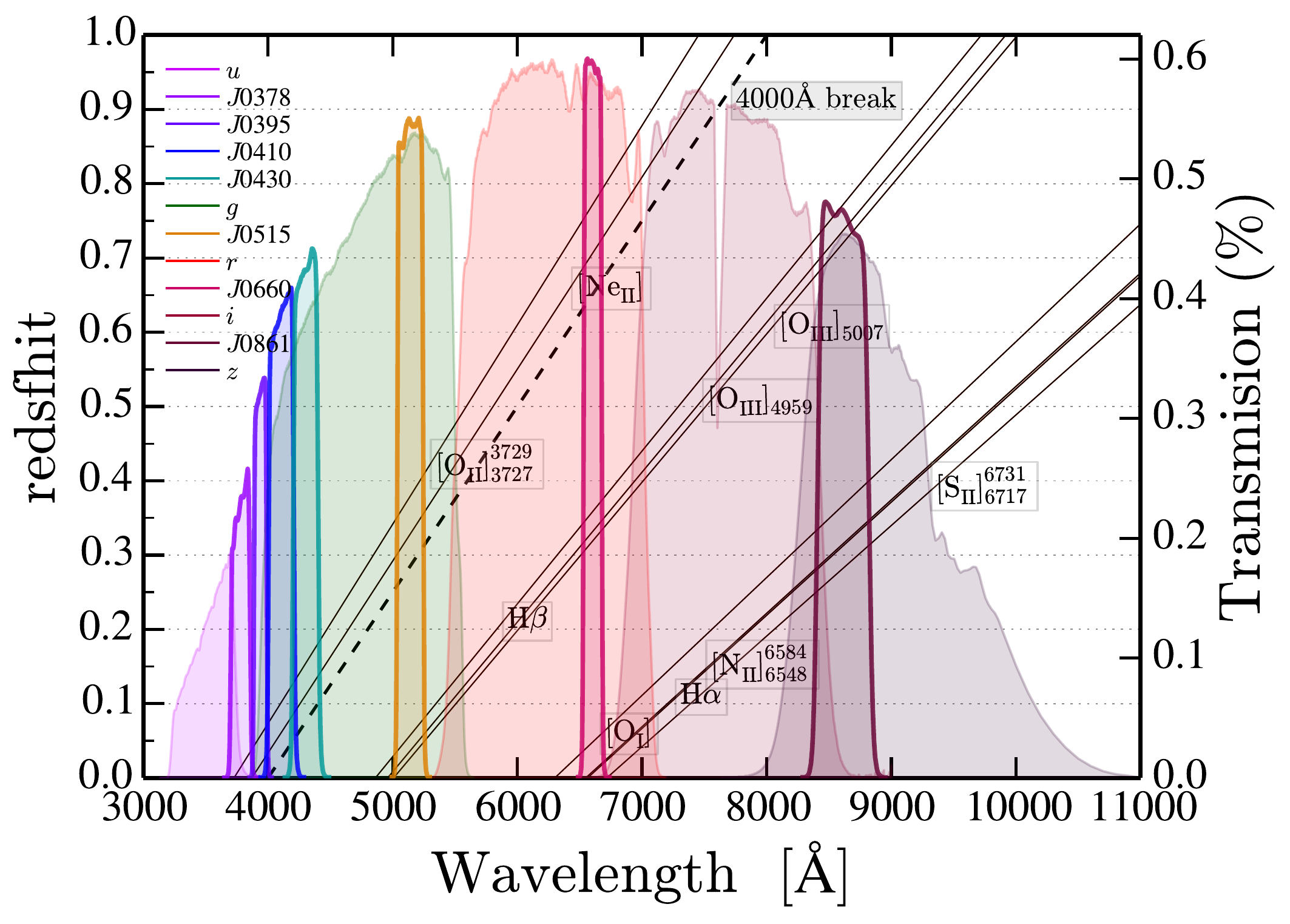}
        \caption{\textit{Right y-axis}: Transmission curves of the J-PLUS system obtained by convolving the measured transmission curves for each filter with the quantum efficiency of the CCD and the atmosphere absorption lines. \textit{Left y-axis}:  Wavelength at which the nine different lines included in the model fall as a function of redshift. The dashed line represents the same, but for the $\rm 4000$ break.}
        \label{fig:JPLUS_filters}
\end{figure}

J-PLUS \citep{Cenarro2018} is an ongoing photometric survey carried out from the Observatorio Astrof\'{i}sico de Javalambre (\textit{OAJ}) in Spain. The J-PLUS collaboration plans to observe thousands of square degrees of the northern sky, of which ${\sim}\,1022\, {\rm deg}^2$ have already been completed and publicly released\footnote{www.j-plus.es/datareleases/data\_release\_dr1} \citep{Cenarro2018}. The survey uses a specially designed camera with a 2 $\rm deg^2$ field of view and 0.55" $\rm pix^{-1}$ scale. The unique feature of J-PLUS is its combination of five broad-band and seven medium-band filters \citep[see Table 3 of][]{Cenarro2018}. We show in Fig.~\ref{fig:JPLUS_filters} the J-PLUS filter transmission curves and the observed wavelengths of nine different lines inside the J-PLUS spectral range as a function of redshift. In this way we can visualise the redshifts at which different emission lines could be selected by various narrow bands. We highlight in red the {\jha} filter ($138\AA$ wide and centred at $6600\AA$), which is expected to capture the \Ha emission of star-forming regions in the nearby universe ($z\,{<}\,0.017$), {\Hb} and {\oiii} at $z\,{\sim}\,0.3$, and {\oii} at $z\,{\sim}\,0.7$, but also the $4000\,\AA$ break at $z\,{\sim}\,0.65$.\\

\begin{figure} 
        \centering
        \includegraphics[width=0.87\columnwidth]{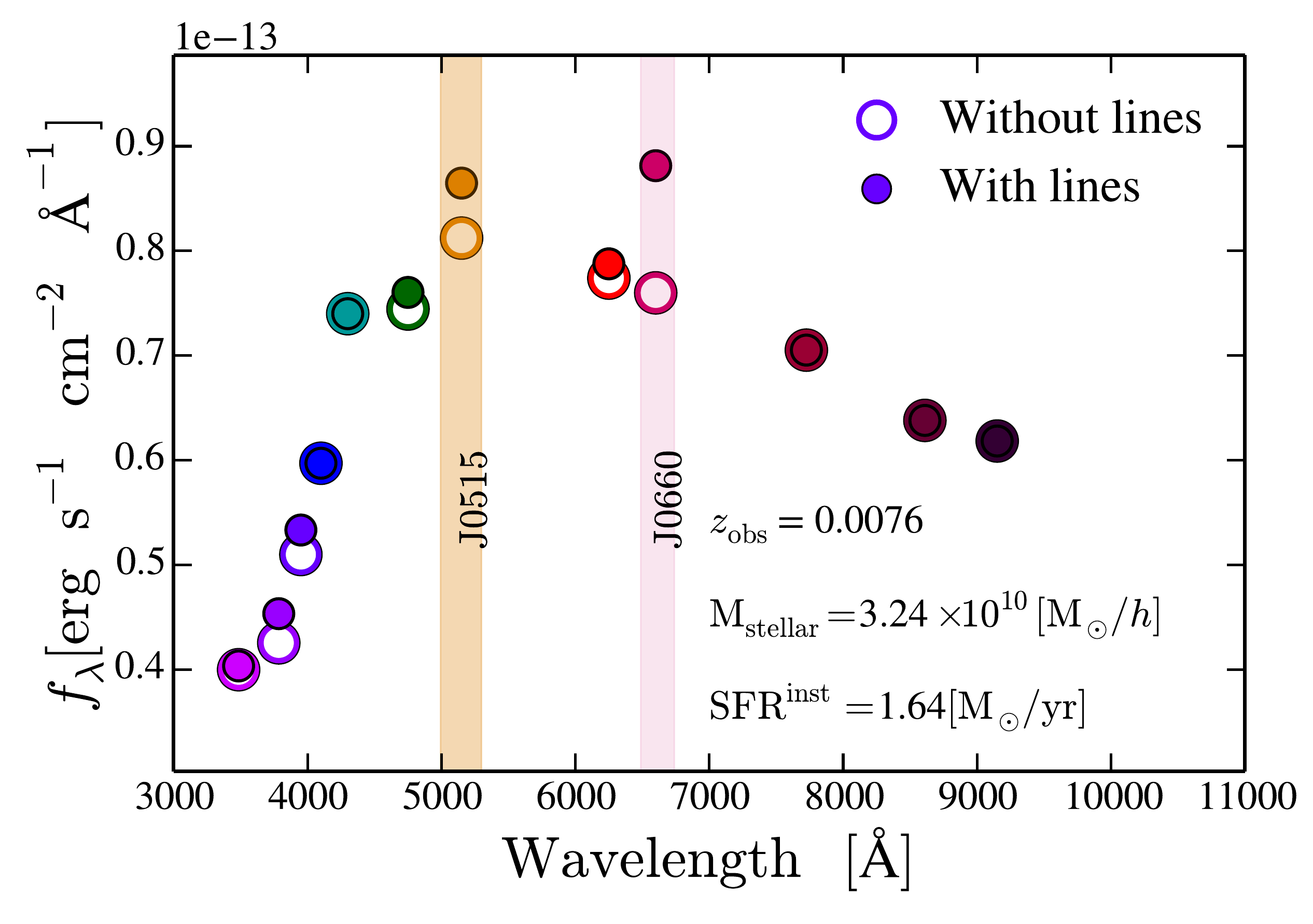}
        \includegraphics[width=0.87\columnwidth]{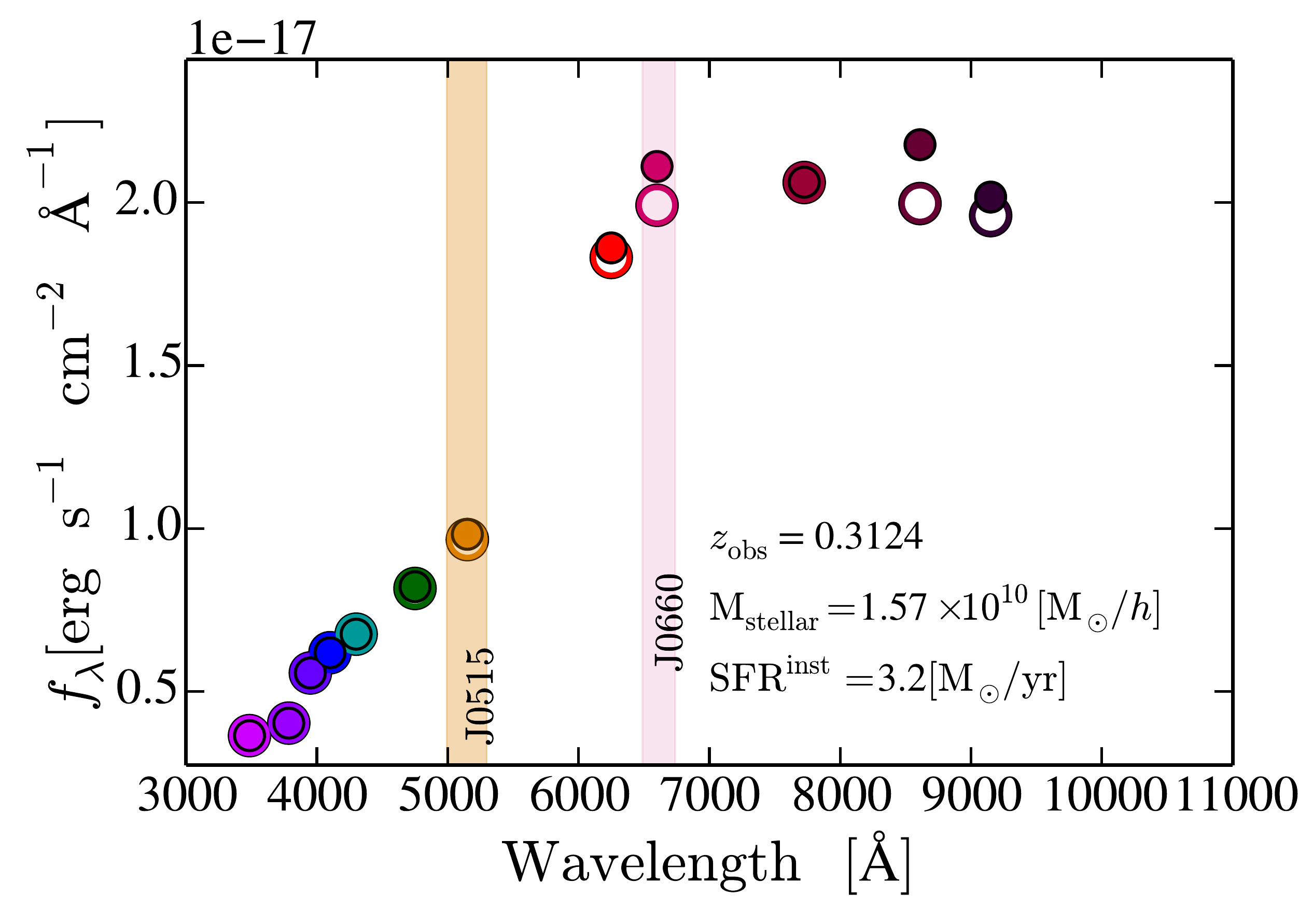}
        \includegraphics[width=0.87\columnwidth]{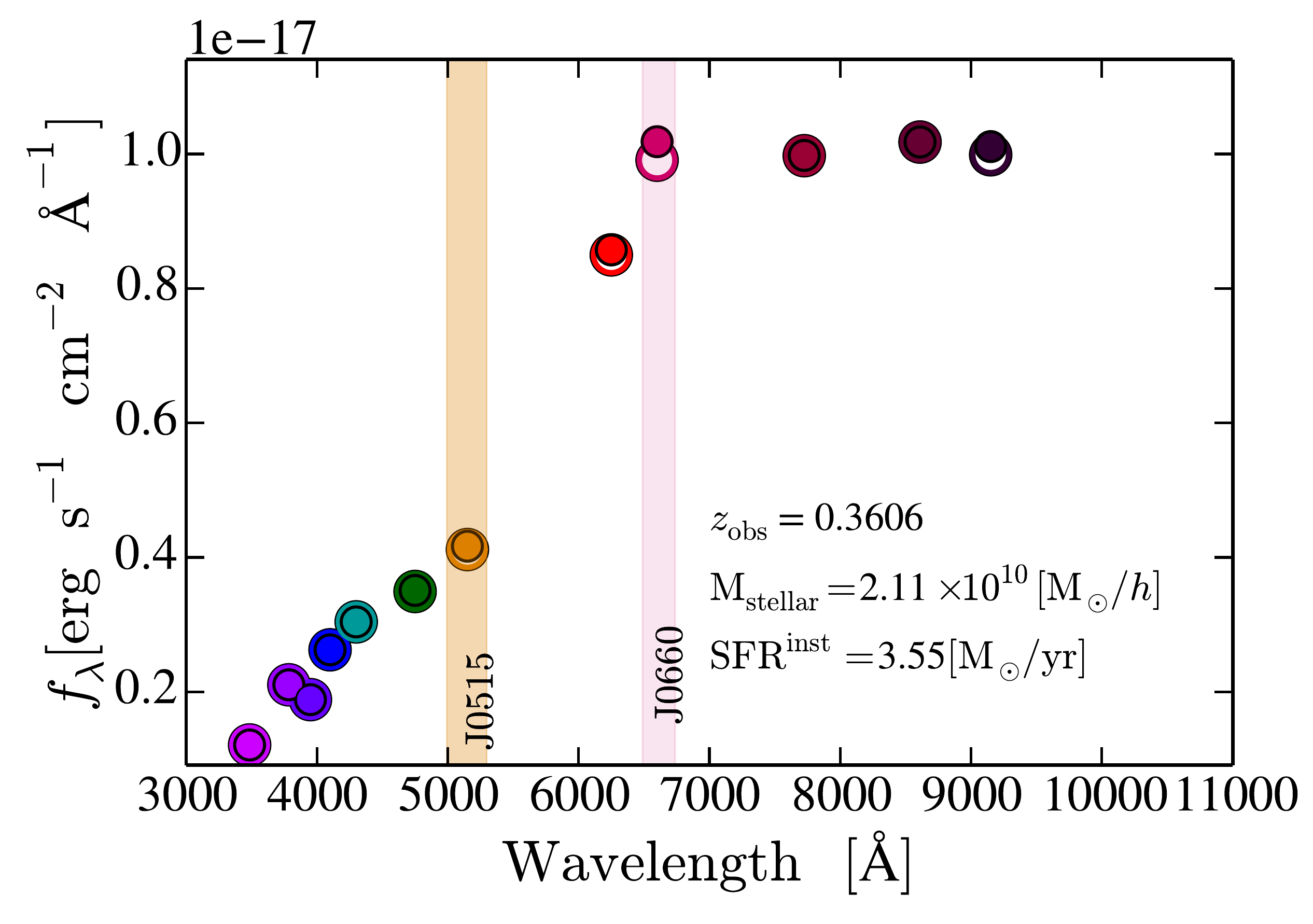}
        \includegraphics[width=0.87\columnwidth]{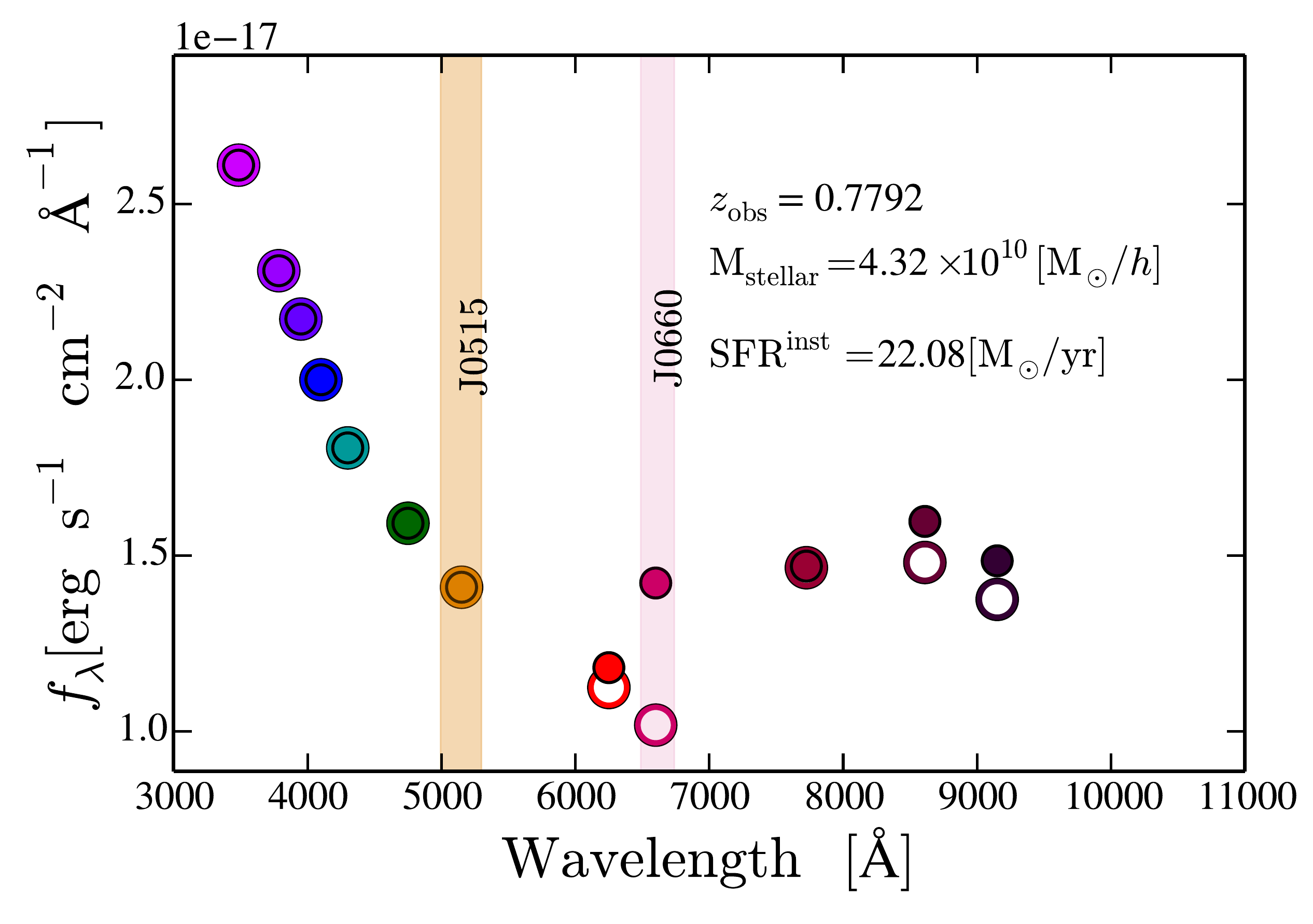}
        \caption{Predicted flux in the 12 J-PLUS filters for four mock galaxies in our lightcone. From left to right: \usdss, $ J\rm0378$, $ J\rm0395$, $ J\rm0410$, $ J\rm0430$, \gsdss, \jOII, \rsdss, \jha, \isdss, $ J\rm0861$, and \zsdss filters. The shaded areas indicate the location and extent of the {\jOII} and {\jha} filters. Empty and filled circles represent the galaxy photometry without and with line contributions, respectively. The first and second panels are examples of \Ha and {\oiii} emitters at $z\,{\sim}\,0$ and $z\,{\sim}\,0.3$, respectively. The third and fourth panels show other examples, but for \Hb and {\oii} emitters at $z\,{\sim}\,0.3$ and $z\,{\sim}\,0.78$, respectively. }
    \label{fig:Galaxy_examples}
\end{figure}

Using the J-PLUS set of transmission curves shown in Fig.~\ref{fig:JPLUS_filters}, and the procedure presented in Sections~\ref{subsection:comp_magnitudes} and \ref{subsec:linecounts}, we  computed synthetic magnitudes for each  galaxy in our mock lightcone. In order to be consistent with the survey, we  kept galaxies with an apparent magnitude \rsdss ${<}\,21.3$, i.e. the $\rm 5\sigma$ detection threshold of galaxies expected in J-PLUS.\\ 

The synthetic J-PLUS photo-spectra for four typical ELGs in our mock is shown in Fig.~\ref{fig:Galaxy_examples}. For each object, we present its photometry including the contribution of emission lines and excluding it. In the first panel of Fig.~\ref{fig:Galaxy_examples} we display a local galaxy ($z_{\rm obs}\,{\sim}\,0$) with a stellar mass of $3.24\,{\times}\,10^{10}\,\Msun$ and an instantaneous star formation rate ($\rm SFR^{inst}$) of $1.64\; \rm M_{\odot}/yr$. We can see how the measured fluxes in the J-PLUS filters are significantly affected by the line emission. Specifically, {\Ha}, {\oii}, and {\oiii} increase  the flux in the filters \jha, $\rm J0378$, and {\jOII}, respectively. Emission lines also affect significantly the broad-band fluxes, as   is the case for {\rsdss} which is affected by \Ha, and for {\gsdss} by {\oiii} (doublet). Thus, the line fluxes have to be taken into account even for broad-band-only analyses. This, for instance, will be important for deep photometric surveys such as LSST.\\

In the second panel of Fig.~\ref{fig:Galaxy_examples}, we show a galaxy at $z\,{\sim}\,0.31$, with stellar mass of $1.57\,{\times}\,10^{10}\,\Msun$ and a ${\rm SFR^{inst}}\,{=}\,3.2\, \rm M_{\odot}/yr$. As in the previous example, the emission lines of this galaxy contribute significantly to the flux measured. However, in this particular case the main line contributing to the {\jha} filter is \oiii, and {\oii} for the {\jOII} filter. The {\Ha} emission is outside the narrow bands, falling in the {\zsdss}-band filter.  In the third panel of Fig.~\ref{fig:Galaxy_examples}, we   show a similar galaxy. In this case the redshift is $z_{\rm obs} \,{\sim}\,0.36$ and the line that falls in the {\jha} is {\Hb}. Finally, in the last panel of Fig.~\ref{fig:Galaxy_examples} we present a high star formation rate galaxy (${\rm SFR^{inst}}\,{\sim}\,22 \, \rm M_{\odot}/yr$) at $z_{\rm obs}\,{\sim}\,0.78$. The main emission lines that can be observed for this galaxy are {\oii} in the narrow-band {\jha} filter and the sum of {\oiii} and \Hb in the $J \rm 0861$ narrow band and \zsdss broad band.\\
The above examples serve as an illustration of the ability of J-PLUS to detect ELGs, but they also show two potential limitations: (i) disentangling the contribution of continuum and emission line to the narrow bands, and (ii) distinguishing the fluxes of different emission lines generated by galaxies at different redshift. 

\subsection{Validation of J-PLUS mocks}

\begin{figure}
        \centering
        \includegraphics[width=\columnwidth]{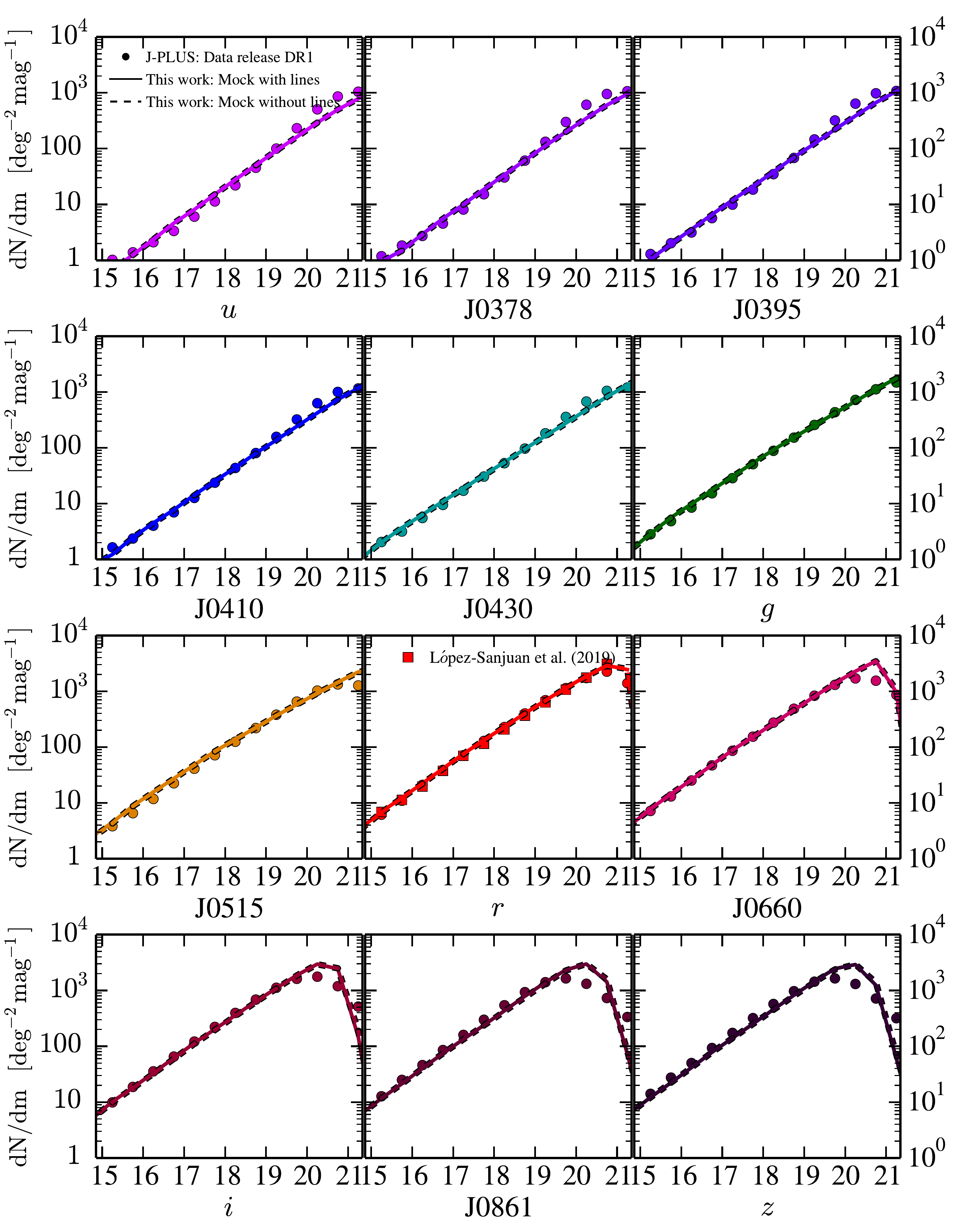}
        \caption{Galaxy number counts in each of the 12 J-PLUS filters. Solid and dashed lines respectively show the mock predictions with and without including the line emission in the galaxy photometry. Symbols show their counterpart in actual J-PLUS observations. The data are from the   DR1 of J-PLUS after masking for saturated objects and applying quality cuts to separate tiles. In the panel of the \rsdss band are included the number counts presented in \cite{LopezSanjuan2019} computed by using the  early data release of J-PLUS.}
        \label{fig:JPLUS_Ncounts}
\end{figure}

\begin{figure}
        \centering
        \includegraphics[width=0.85\columnwidth]{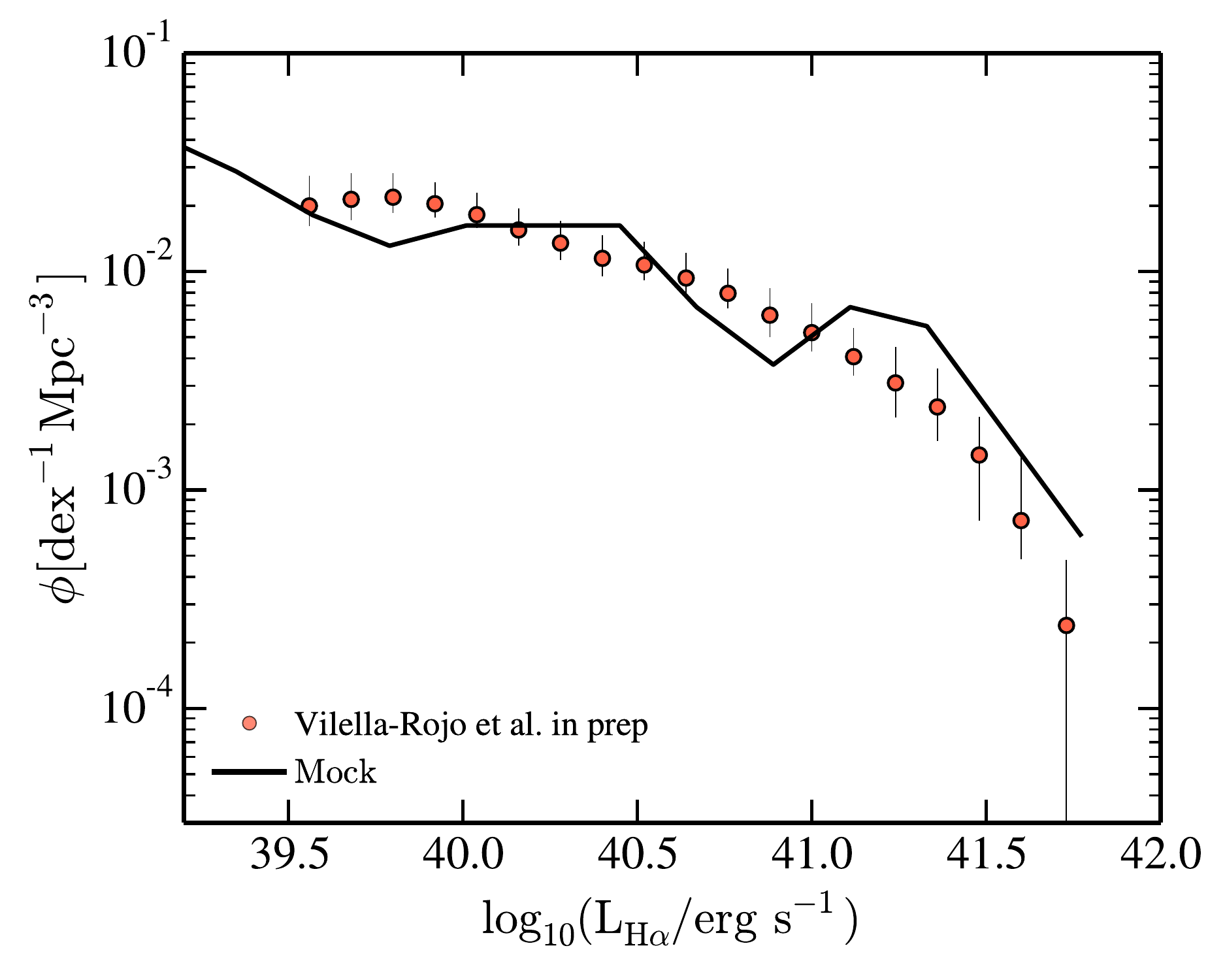}
        \caption{Intrinsic {\Ha} luminosity function of the local universe ($z\,{<}\,0.017$). Red symbols indicate the observational results presented in Vilella-Rojo et al. (2019, in prep.) using J-PLUS. The solid black line displays the predictions from our J-PLUS mock catalogue.}
        \label{fig:LF_Ha_JPLUS}
\end{figure}
In this section we further validate our mock by showing the  agreement with the currently available data of the survey \citep[][]{Cenarro2018}.\\


Firstly, the number counts of the 12 J-PLUS bands are presented in Fig.~\ref{fig:JPLUS_Ncounts}. Galaxies in the J-PLUS  Data Release 1 (DR1) have been selected by imposing the morphological star--galaxy classification parameter of \cite{LopezSanjuan2019} to be less than 0.5. The match between mocks and observations is remarkable. We note that the agreement in the bluest narrow bands starts to fail at magnitude ${\gtrsim}\,19.5$. This is principally because our SAM variant underestimates the population of blue counts (see Section~\ref{sec:validation}). In the same plot, we   added   the number counts after removing the line contribution in the galaxy photometry. We do not see significant differences for the global population of galaxies. This is expected since only a small fraction of galaxies would display emission lines falling within one of the J-PLUS narrow bands.\\


Finally, in Fig.~\ref{fig:LF_Ha_JPLUS} we present the predicted and observed {\Ha} luminosity function in the local universe ($z\,{<}\,0.017$) seen by the J-PLUS survey. The observational results can be found in Vilella-Rojo et al. (in prep.). Again, predictions and observations agree with each other. The fact that our LF is not as smooth as the observed one is due to the limitation imposed by the cosmic variance at such low-$z$. This is the result of the narrow angular aperture that characterises our lightcone. 

\subsection{Selecting emission-line galaxies} \label{sec:EL_Extraction}

We now use the 3FM method developed by \cite{VilellaRojo2015} to estimate the emission-line flux from a linear combination of broad- and narrow-band filters. In short, this method infers the continuum of galaxy in a narrow band (\jha) by linearly interpolating the continuum using two adjacent broad-band filters (\rsdss and \isdss). We note that this method takes into account the emission line contribution in the broad band when performing the interpolation. Using synthetic photometry computed from SDSS spectra, \cite{VilellaRojo2015} demonstrated that for $z\,{\sim}\,0$ galaxies the method is nearly unbiased ($\lesssim 9$\%) in extracting \Ha emission.\\

In the following we explore higher redshifts ($z\,{>}\,0.017$, implying no \Ha emission in the {\jha} filter) and we asses the performance of the 3FM method to extract line emission of high redshift galaxies. To this end, we  applied the 3FM method to every galaxy in our J-PLUS mock using the {\jha} narrow band as a line tracer and the  \rsdss and \isdss broad-band filters to estimate the galaxy continuum, $\rm  m_{J0660}^{cont,Est}$. We built the magnitude excess, $\rm \Delta m$, as follows:
\begin{equation}
\rm \Delta m \,{=}\, m_{J0660}^{cont,Est} - m_{J0660}.
\end{equation}
\noindent Here $\rm m^{J0660}$ is the observed magnitude in the \jha. We selected objects with an excess of flux such that $\rm \Delta m \,{>}\,0$. This is close to imposing an equivalent width (EW)\footnote{The equivalent width is defined as the ratio of the total line flux, $\rm F(\lambda_j|q,Z_{cold})$, to the continuum  density flux, $f_{\lambda}^{c}$, at the line position.} cut on the galaxy line emission\footnote{By assuming a $\delta$-Dirac line profile we can establish the relation  ${\rm \Delta m = 2.5\log_{10}\left[ 1 + \left( \lambda_{Line}^{obs} \, T(\lambda_{Line}^{obs})/\int{\rm \lambda \, T(\lambda)\, \mathit{d}\lambda}\right) EW\right]}$, where $\rm \lambda_{Line}^{obs}$ is the observed line wavelength, $\rm T(\lambda)$ is the narrow-band filter transmission curve, and $\rm EW$ is the line equivalent width.}.\\

\begin{figure} 
        \centering
        \includegraphics[width=\columnwidth]{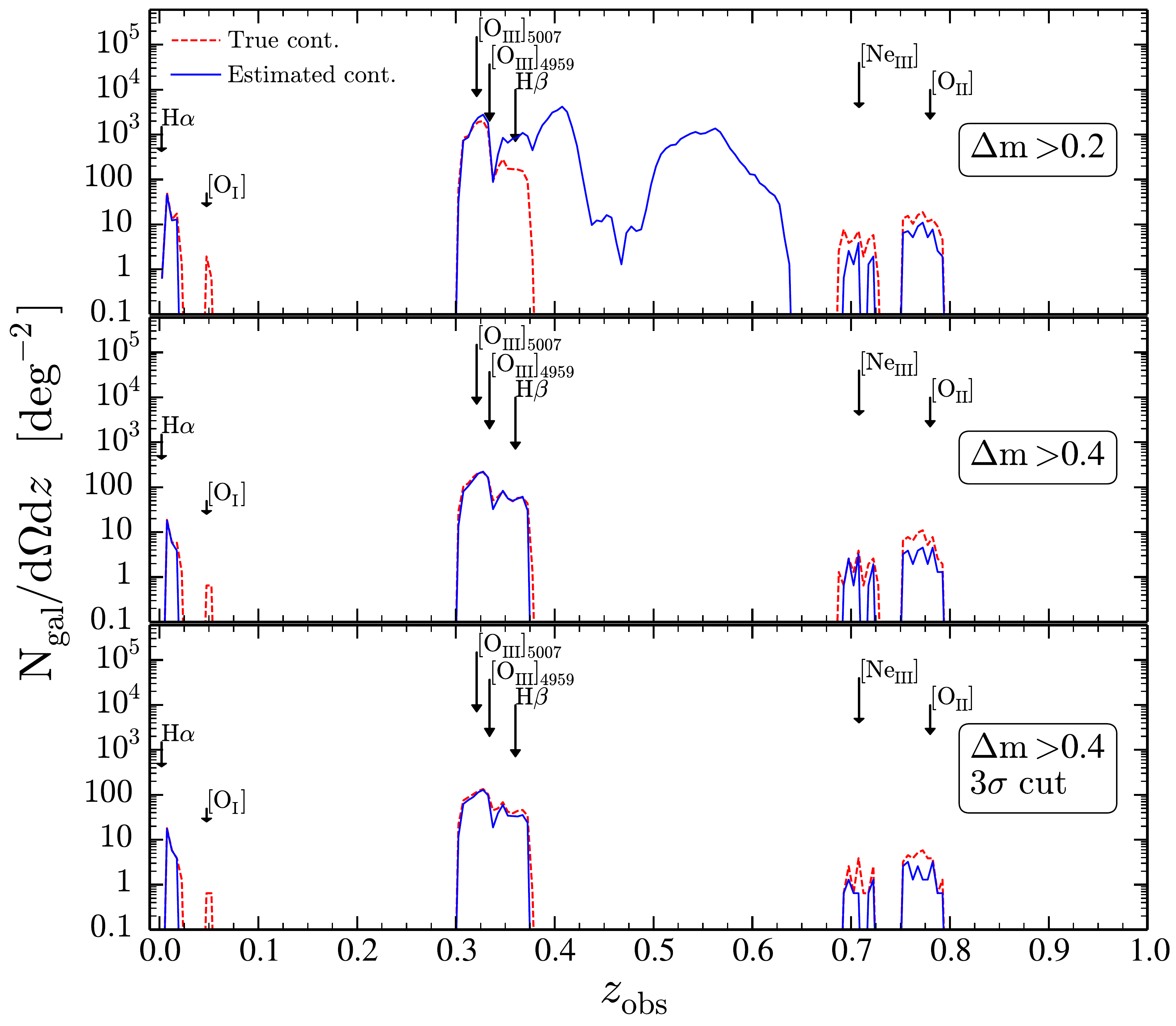}
        \caption{Redshift distribution of galaxies selected to have a positive excess
                in the {\jha} filter associated to line emission. The top, middle, and bottom panels display
the                results employing different magnitude thresholds, $\rm \Delta m\,{>}\,0.2$, 
                $\rm \Delta m\,{>}\,0.4$, and $\rm \Delta m\,{>}\,0.4$, plus the $3 \sigma$ significance level with respect to the J-PLUS typical uncertainties. The blue histograms show galaxies selected using the 3FM of \cite{VilellaRojo2015},      whereas the red dashed histograms show galaxies selected using the true magnitude excess associated with a line emission.}
        \label{fig:redshift_distribution}
\end{figure}

We present the redshift distribution of selected galaxies in Fig.~\ref{fig:redshift_distribution}. 
From left to right, the red peaks in the distribution correspond to \Ha, \oi, {\oiii} (4959\AA, 5007\AA), \Hb, $\rm [\ion{Ne}{II}],$  and the \oii \space doublet at redshift $0.01$, $0.05$, $0.33$, $0.36$, $0.7$, and $0.78$, respectively. When using a minimum detection of $ \rm \Delta m\,{>}\,0.2$, we see that most of the objects selected  correspond to galaxies at $z\,{\in}\,[0.3,0.7]$ that do not necessarily present a significant line emission. This implies that a sample selected using this threshold is contaminated by a significant fraction of spurious detections. As we  see in Section~\ref{sec:interlopers}, these spurious detections are produced by the non-linear behaviour of the continuum for the concerned range of wavelength, which corresponds to the 4000$\AA$ break crossing the J-PLUS set of filters {\rsdss}, {\jha}, and {\isdss} in the range $0.3 \,{\lesssim}\,z\,{\lesssim}0.8$.\\

When we apply a  higher threshold,  $\rm \Delta m\,{>}\,0.4$, the spurious selection is significantly reduced and the detections correspond to galaxies with emission lines, covering the correct range of redshift. For this $\rm \Delta m$ threshold,  74.7\% of the selected emitters are \oiii, 13.1\% \Hb, 10.4\% \Hb + \oiii\footnote{The \Hb and {\oiii} overlap inside {\jha} in the redshift range $0.34\,{\lesssim}\,z\,{\lesssim}\,0.36$.}, 0.9\% \Ha, 0.2\% $\rm [\ion{Ne}{II}]$, and 0.7\% \oii. We note that with the 3FM method we are not able to detect the \oi \space emitters due to the weak emission of this line. In addition, independently of the $\rm \Delta m$ cut, the  $\rm [NeII]$ and \oii \space emitters have a low completeness; in other words,  the true distribution (dashed red line) is above the recovered distribution (solid blue line).\\

Finally, to mimic a more realistic scenario, in the bottom panel of Fig.~\ref{fig:redshift_distribution} we also account for the photometric uncertainties of the J-PLUS survey. For this, we applied an extra cut such that the magnitude excess in the narrow-band filter is above a significance of 3$\sigma$ with respect to the J-PLUS typical uncertainties. To estimate the J-PLUS data uncertainty level, we used the median uncertainty reported in the J-PLUS   DR1 as a function of {\rsdss} magnitude. As we can see, the 3$\sigma$ significance cut leaves the distribution almost unchanged  when we compare it with the distribution that does not have such cut (middle panel). The high-$z$ emitters are the most affected by this extra cut as they are faint sources with typically larger photometric uncertainties.

\begin{figure}
        \centering
        \includegraphics[width=\columnwidth]{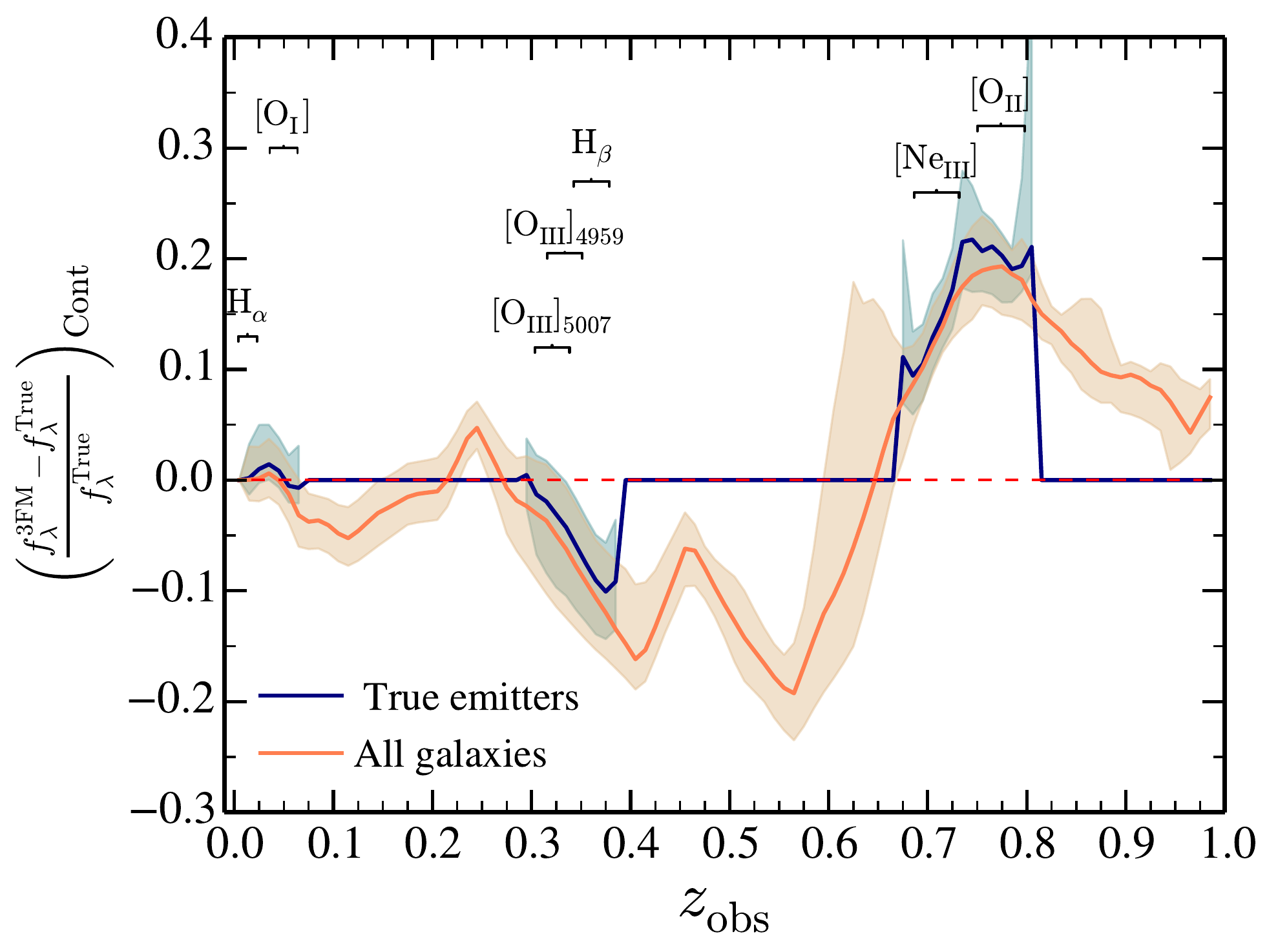}
        \caption{Comparison between the true, $f_{\lambda}^{\rm True}$, and inferred, $f_{\lambda}^{\rm 3FM}$, values for the continuum density flux. The orange line represents the comparison for all the galaxies (with and without emission lines) in the mock. The blue line are the same, but only for galaxies displaying emission line features in the filter \jha. In both cases, the shaded areas enclose the 25th and 75th percentiles.}
        \label{fig:Deltas}
\end{figure}

\subsection{Understanding the population of interlopers} \label{sec:interlopers}

To understand the origin of the \textit{interlopers} and the under-recovered population of $\rm [NeII]$ and \oii \space in Fig.~\ref{fig:Deltas}, we compare the true, $f_{\lambda}^{\rm True}$, and inferred, $f_{\lambda}^{\rm 3FM}$, values for the continuum density flux. From Fig.~\ref{fig:Deltas} we confirm the findings of \cite{VilellaRojo2015}, in that the 3FM provides a nearly unbiased estimate of the continuum at $z\,{\sim}\,0$. However, we find significant biases at higher redshifts,  most notably at $z\,{\sim}\,0.4$ and $0.6$ where the continuum is overestimated by ${\sim}\,20 \%$.\\ 

\begin{figure} 
        \centering
        \includegraphics[width=0.45\textwidth]{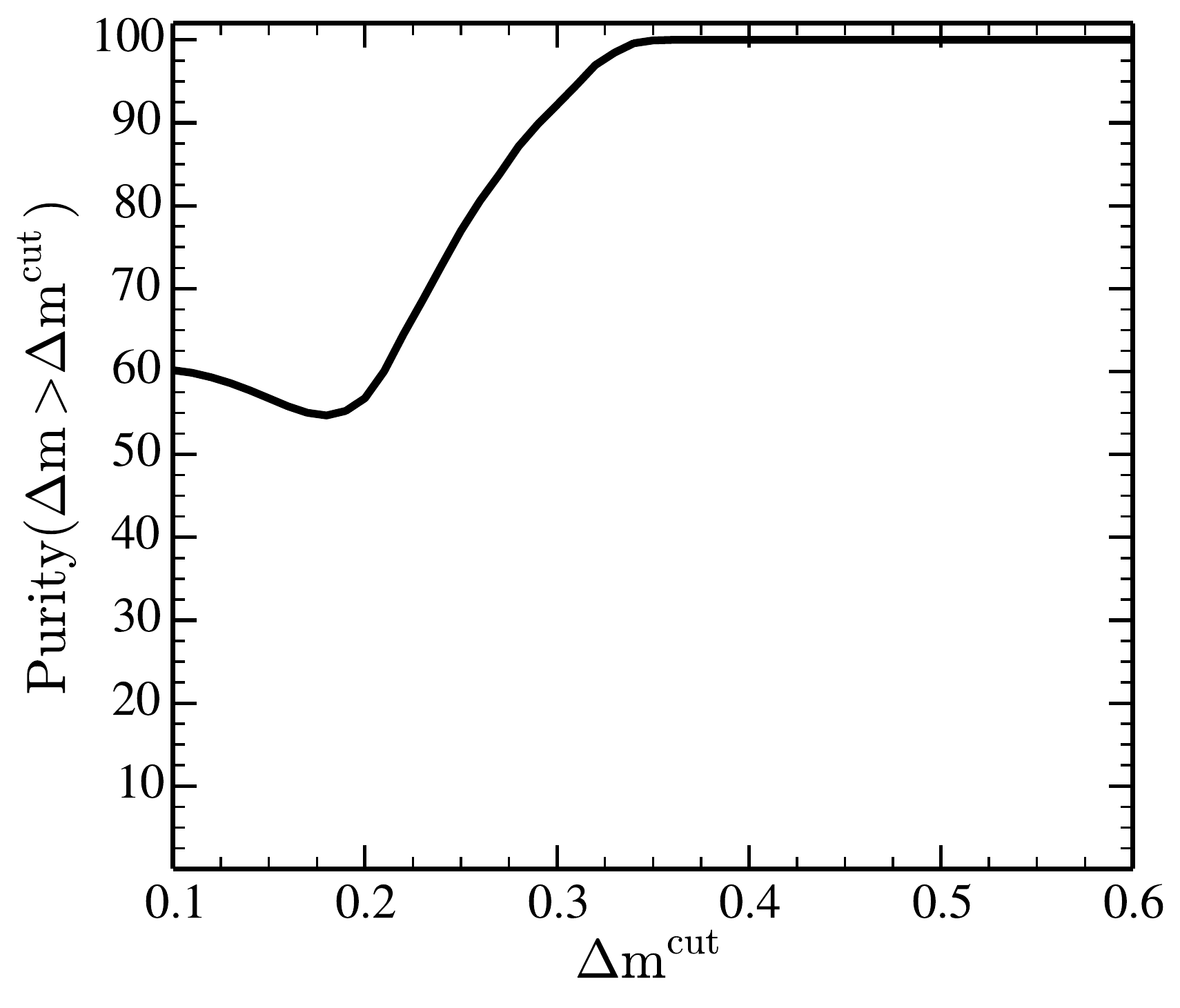}
        \caption{Purity in a catalogue of mock J-PLUS galaxies selected to have emission lines, as a function of the threshold used for detection $\rm \Delta m^{cut}$. }
        \label{fig:Purity_and_comp}
\end{figure}

Moreover, there is an underestimation of the continuum at $z\,{\sim}\,0.8$ of about $20 \%$ \footnote{We  computed this bias as a function of different \rsdss apparent magnitudes [18,19], [19,20], and [20,21], and found almost identical results.}. A similar behaviour can be found for the true emission-line galaxies (blue line in the figure). To investigate the origin of these trends, we  applied the 3FM to magnitudes that exclude the contribution of emission lines. We found that the trends are preserved, indicating that the biases are caused by features in the spectral energy distribution (SED) continuum of galaxies. We expect any non-linear feature in a galaxy spectrum to produce biases in the continuum estimation. In particular, the 4000$\AA$ break crossing our set of filters \rsdss, {\jha} and {\isdss} between $0.40\,{\lesssim}\, z \,{\lesssim}\,0.8$ is responsible for the systematic overestimation of the continuum at $z\,{\sim}\,0.6$ and the underestimation at $z\,{\sim}\,0.8$. The typical curvature of our mock galaxy photo-spectra moves from positive to negative. A correction of this continuum subtraction bias could be applied by combining photo-$z$ information with a flux correction based on the orange curve of Fig.~\ref{fig:Deltas}. However, this procedure goes beyond the scope of this paper.\\

The purity of ELGs as a function of magnitude excess cut is presented in Fig.~\ref{fig:Purity_and_comp}. The curve displays a decreasing trend between $\rm 0.1\,{<}\, \Delta m^{cut} \,{<}\,0.2$, a consequence of the fact that these $\rm  \Delta m$ cuts remove low equivalent width ELGs, yet   still keep the majority of \textit{interlopers}. As soon as $\rm \Delta m^{cut}\,{\gtrsim}\,0.2$ we start to avoid \textit{interlopers}, recovering the increasing trend. As shown, the 4000 {\AA} break is not capable of generating a fake magnitude excess above $\rm \Delta m^{cut}{\sim}0.36$, where we obtain a purity of $100\%$.\\

\begin{figure} 
        \centering
        \includegraphics[width=1.0\columnwidth]{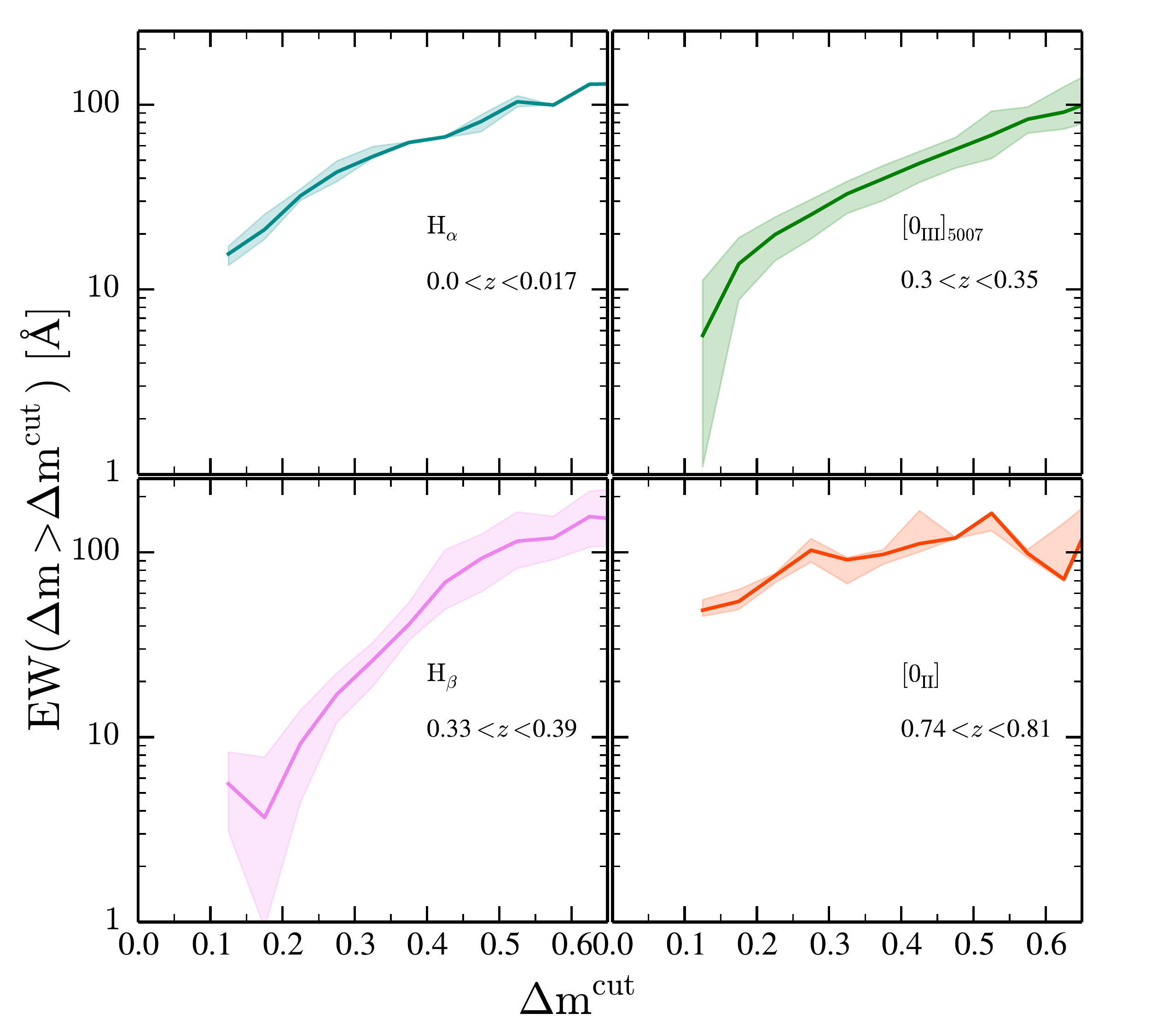}
        \caption{Relation between $\rm \Delta m^{cut}$ and the equivalent width of the line ($\rm EW$). From upper left to lower right: \Ha at $0\,{<}\,z\,{<}\,0.017$,  {\oiii}  at $0.3\,{<}\,z\,{<}\,0.35$, \Hb at $0.33\,{<}\,z\,{<}\,0.39$, and {\oii} at $0.74\,{<}\,z\,{<}\,0.81$. The solid lines represent the median EW for the (emitters) sample with $\rm \Delta m\,{>}\,\Delta m^{cut}$; the shaded areas represent the 1$\sigma$ value of the distribution.} 
        \label{fig:EW_Delta_m_Est_relation}
\end{figure}

Finally, in Fig.~\ref{fig:EW_Delta_m_Est_relation} we present the typical equivalent width as a function of $\rm \Delta m$ cuts for the different emission lines that contribute in the {\jha} filter i.e. \Ha at $0\,{<}\,z\,{<}\,0.02$, {\oiiiFd}  at $0.3\,{<}\,z\,{<}\,0.35$, \Hb at $0.33\,{<}\,z\,{<}\,0.39$ and {\oii} at $0.74\,{<}\,z\,{<}\,0.81$. We find that while $\rm \Delta m\,{\lesssim}\,0.2$ imposes {\Ha}, {\Hb}, and {\oiiiFd} EW cuts of ${\sim}10\,\AA$, it implies a much more strict cut for {\oii} with an $\rm EW \,{\sim}\,50\,\AA$. When we increase $\rm \Delta m^{\rm cut}$, we can see that a more severe $\rm EW$ requirement is imposed for all the lines. In particular, $\rm \Delta m\,{>}\,0.4$ implies $\rm EW\,{>}\,70\,\AA$.\\

All this points towards the good capability of J-PLUS to study ELGs in the universe. The forthcoming J-PAS survey \citep{Benitez2014} will increase the capabilities of detecting line-emission galaxies due to the higher number of narrow bands (56) and the higher depth, with respect to J-PLUS. The methods we developed here can be easily generalised to the J-PAS case, and can thus provide a efficient tool for testing the survey's data capabilities.

\section{Summary and conclusions}
\label{sec:conclusions}

In this paper we have presented a new procedure to generate synthetic galaxy lightcones specifically designed for narrow-band photometric surveys. Different from previous lightcone construction methods, we  embedded its assembly inside the galaxy formation modelling so that each galaxy is evolved up to the exact moment it crosses the past lightcone of a given observer. This produces accurate results across cosmic time, while minimising time-discreteness effects. Specifically, we  used \lgs \citep{Guo2011} implemented on top of the dark matter merger trees of the \texttt{Millennium} $N$-body simulation \citep{Springel2005}.
Since the \texttt{Millennium} box size is not able to cover the whole survey volume, we  replicated the box eight times in each spatial direction, corresponding to a maximum redshift $z\,{\sim}\,3$, large enough to include high-$z$ ELGs. With the purpose of minimising the repetition of large-scale structures, we  placed the observer in the origin of the first replication with a LOS orientation of $(\rm \theta, \varphi) \,{=}\,(58.9\degree , 56.3\degree)$. The angular extent of the lightcone was chosen to be $\rm {22.5\degree}\,{\times}\,{22.5\degree}$, i.e. no more than two repetitions of the simulation box would be required to represent the cosmic structure up to $z\,{\sim}\,1.0$.\\

As a particular feature of our mock, we  included the effect of nine different emission lines in the final galaxy photometry. In particular, $\rm Ly_{\alpha}$(1216\AA), \Hb (4861\AA), \Ha (6563\AA), \oii \space (3727\AA, 3729\AA), $\rm [\ion{Ne}{III}]$ (3870\AA), {\oiii} (4959\AA, 5007\AA), {\oi} (6300\AA), $\rm [\ion{N}{II}]$ (6548\AA, 6583\AA), and $\rm [\ion{S}{II}]$ (6717\AA, 6731\AA). This is one of the first times that multiple emission lines have been included in  mock galaxy cones \citep{Merson2018,Stothert2018}. The properties of these lines were  computed using the \cite{Orsi2014} model for nebular emission from star-forming regions. Based on \texttt{MAPPINGS-III} photo-ionisation code the model predicts different line luminosities according to the galaxy gas metallicity, instantaneous star formation rate, and ionisation parameter. For the two former quantities  we  used the predictions of our mock SAM galaxies.\\


We presented various tests to validate our lightcone construction. Galaxy photometry has been tested with the galaxy number counts in the \usdss, \gsdss, \rsdss, \isdss, {\zsdss} broad bands. In the case of galaxy spatial distribution we  compared the clustering of \gsdss selected galaxies with the work of \cite{Favole2016}. In both cases the agreement is good. By comparing our mock line-luminosity functions to observational works we calibrated our dust attenuation. It was based on a dependence with the galaxy redshift and metallicity. In particular, we compared our results with to the well-constrained \Ha, \Hb, {\oii}, and {\oiiiFd} luminosity functions to develop a global line attenuation for all the lines included in our mock. Despite its limitations, this simple method produces final luminosity functions in good agreement with observations, even with respect to the observed redshift evolution.\\



As an application of our lightcone, we have generated catalogues tailored to the photometry of the ongoing J-PLUS survey \citep{Cenarro2018}. With these mocks we have studied the ability of the survey to correctly identify emission-line galaxies at various redshifts. In particular, among all the intermediate- and narrow-band filter available to detect lines, we have focused in the {\jha} which is able to capture the \Ha emission of star-forming regions in the nearby universe ($z\,{<}\,0.017$) and other lines at higher redshifts ($z\,{>}\,0.3$) such as \Hb, {\oiii,} and {\oii}. To assert the detection of the emission lines in J-PLUS, we used the \textit{three-filters method} developed by \cite{VilellaRojo2015}. Our mocks proved that the extraction of  emission lines is strongly dependent on the continuum shape. In particular, we showed that the $4000\AA$ break in the spectral energy distribution of galaxies can be misidentified as line emission, selecting a population of \textit{fake} emission-line galaxies at $0.3\,{<}\,z\,{<}\,0.6$. However, we showed that all significant excess in the narrow band (larger than 0.4 magnitudes) can be correctly and unambiguously attributed to emission-line galaxies. The mock catalogue is publicly available at  \href{https://www.j-plus.es/ancillarydata/mock\_galaxy\_lightcone}{https://www.j-plus.es/ancillarydata/mock\_galaxy\_lightcone}.\\

In summary, in this work we have presented a new approach used to mimic photometric narrow-band survey observations. We have shown that the synergy between galaxy formation models, dark matter $N$-body simulations, and photo-ionisation codes is an adequate combination for the creation of realistic mocks for the next generation of narrow-band photometric surveys. In addition, we anticipate that our work will be an important tool for  correctly interpreting narrow-band surveys and for quantifying the impact of line emission in broad-band photometry. As a future application the procedure presented here would be extended to the J-PAS survey \cite{Benitez2014} whose unique feature of 56 narrow-band filters would require mock galaxy catalogues to exploit its data capabilities.

\section*{Acknowledgements}

The authors contributed as follows to this paper: DIV carried out the majority of the work presented. RA and AO supervised DIV and helped with the SAM and the emission-line modelling. GH, GVR, SB, and CLS provided interesting inputs to the paper and helped with interpretation of the results. The rest of the authors contributed to developing the J-PLUS survey.
\vspace{2mm}

The authors thank Elmo Tempel, Roderik Overzier, David Sobral, and Luis A. Díaz-García for useful comments. DIV acknowledges the grant \textit{Programa Operativo Fondo Social Europeo de Arag\'{o}n 2014-2020. Construyendo Europa desde Arag\'{o}n}. REA acknowledges the support from the European Research Council through grant number ERC-StG/716151. DIV particularly thanks Daniele Spinoso for helping with the J-PLUS data and for the interesting discussions.  Thanks are due to Ginevra Favole for kindly providing her observational results and to Tamara Civera for developing the J-PLUS mock web page. Funding for the J-PLUS Project has been provided by the Governments of Spain and Arag\'on through the Fondo de Inversiones de Teruel; the Arag\'on Government through the Reseach Groups E96, E103, and E16\_17R; the Spanish Ministry of Economy and Competitiveness (MINECO; under grants AYA2015-66211-C2-1-P, AYA2015-66211-C2-2, AYA2012-30789 and ICTS-2009-14); and  European FEDER funding (FCDD10-4E-867, FCDD13-4E-2685). 

\vspace{0.5cm}

\bibliographystyle{aa} 
\bibliography{Ref}
\appendix

\section{Minimum structure repetition} \label{appendix:repetition}

\begin{figure}[h]
        \centering
        \includegraphics[width=\columnwidth]{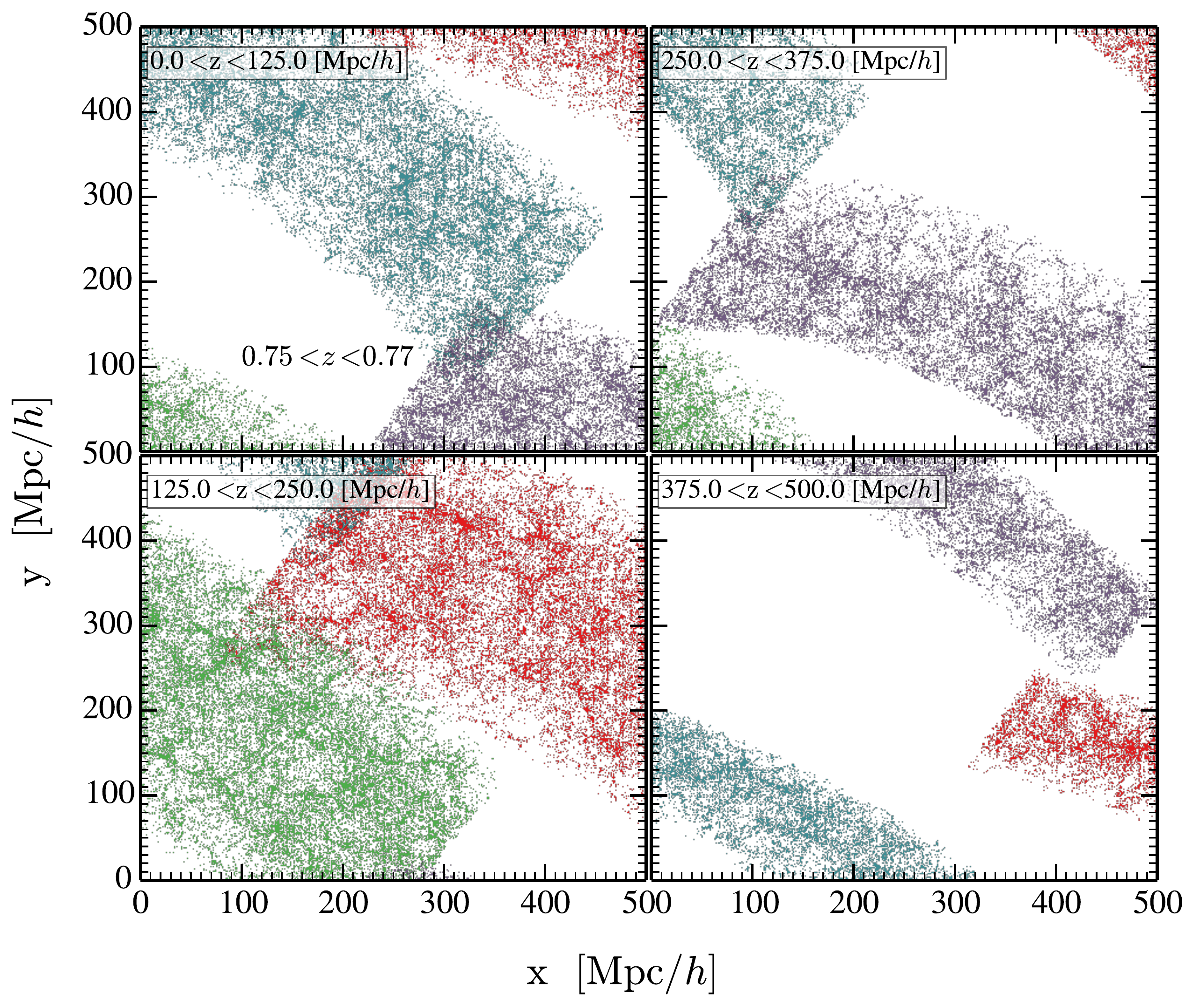}
        \caption{Example of the minimum repetition between \texttt{Millennium} box replications. For four   different z-axis thicknesses  the plane x-y is shown for galaxies in the redshift bin $0.75\,{<}\,z\,{<}\,0.77$. To check the structure repetition   the modulus 500 Mpc/$h$ (box size) of the x and y position was used. Each colour represents a different box replication. A minimum overlap is present.}
        \label{fig:solap}
\end{figure}

In Fig.~\ref{fig:solap} we present, for different z-axis slabs, the original (x,y) coordinates (i.e. without replication) for galaxies in the redshift range $0.75\,{<}\,z\,{<}\,0.77$. Each  colour represents a different box replication. As we can see, the overlap between the same structures belonging to different replications boxes is minimum. The bigger the redshift range, the larger the overlapping will be.\\

\section{Luminosity function evolution} \label{appendix:LF}
\begin{figure*}
        \centering
        \includegraphics[width=1.75\columnwidth]{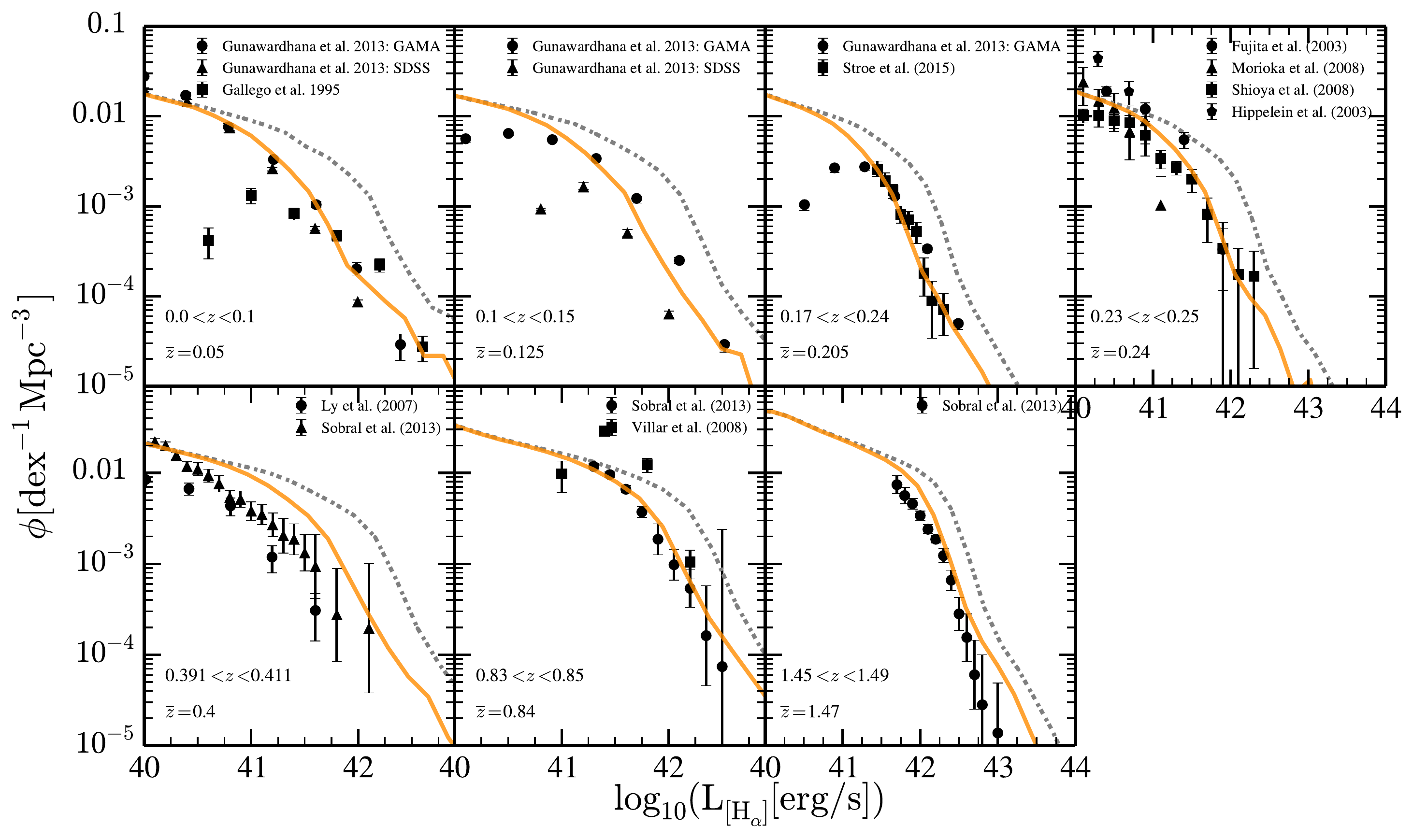}
        \caption{Luminosity function of {\Ha} line at seven different redshifts. Shown are the  comparisons with the \protect{\cite{Gallego1995}},\protect{\cite{Fujita2003}}, \protect{\cite{Gunawardhana2013}}, \protect{\cite{Sobral2013}} and \protect{\cite{Stroe2015}} observational data.}
        \label{fig:ELG_LF_Halpha}
\end{figure*}

\begin{figure*}
        \centering
        \includegraphics[width=1.75\columnwidth]{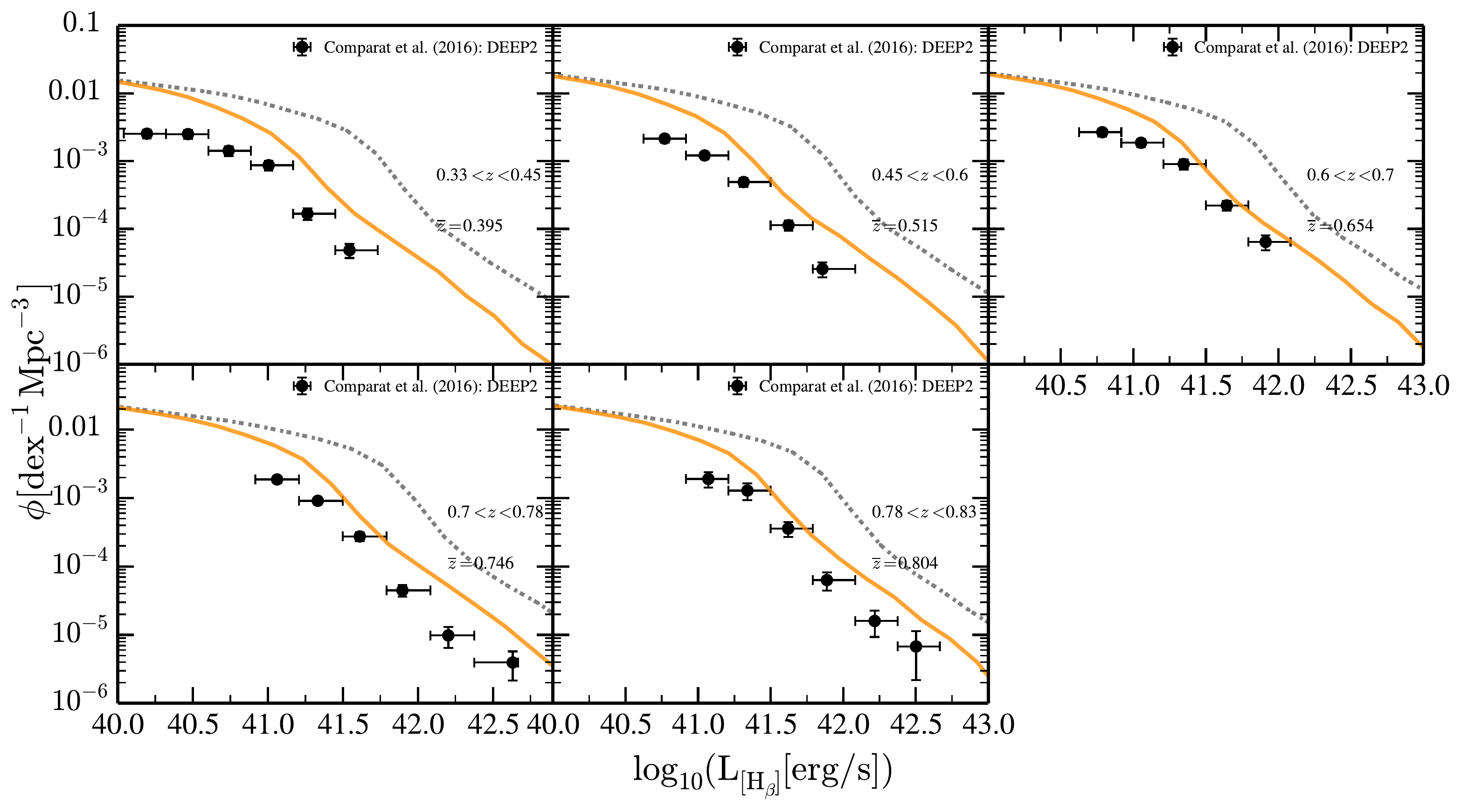}
        \caption{Luminosity function of {\Hb} line at five different redshifts. Shown is the  comparison with the recent observational work of \protect{\cite{Comparat2016}}. The black dots represent the observational data, while the solid orange line and grey dashed line the predictions of our mocks LF with and without dust attenuation.}
        \label{fig:ELG_LF_Hb}
\end{figure*}

\begin{figure*}
        \centering
        \includegraphics[width=1.75\columnwidth]{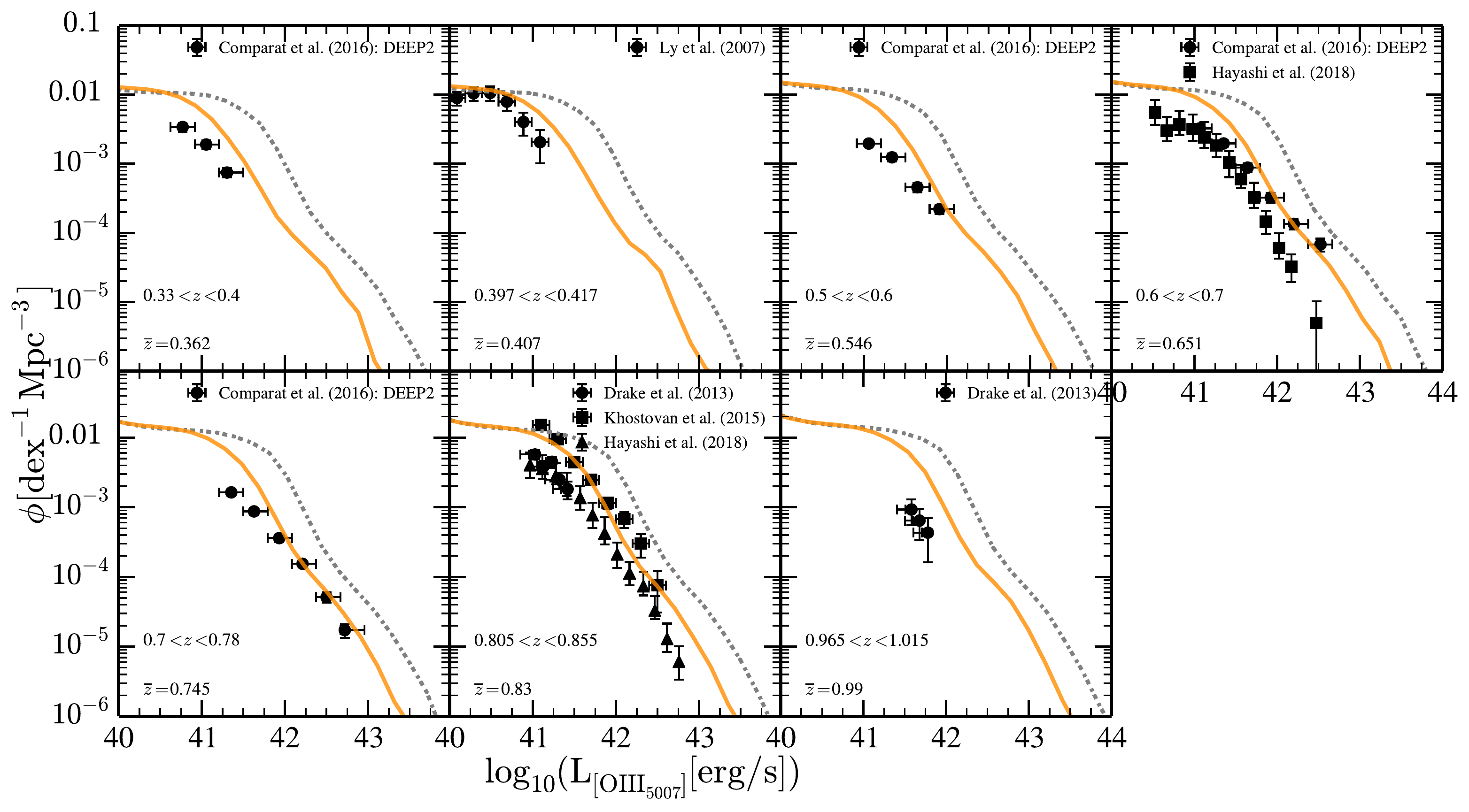}
        \caption{Luminosity function of {\oiiiFd} line at seven different redshifts. Shown are the  comparisons  with the recent observational work of \protect{\cite{Comparat2016}},\protect{\cite{Ly2007}}, \protect{\cite{Drake2013}}, \protect{\cite{Khostovan2015}} and \protect{\cite{Hayashi2018}}. The black dots represent the observational data, while the solid orange line and grey dashed line the predictions of our mocks LF with and without dust attenuation.}
        \label{fig:ELG_LF_OIII}
\end{figure*}

\begin{figure*}
        \centering
        \includegraphics[width=1.75\columnwidth]{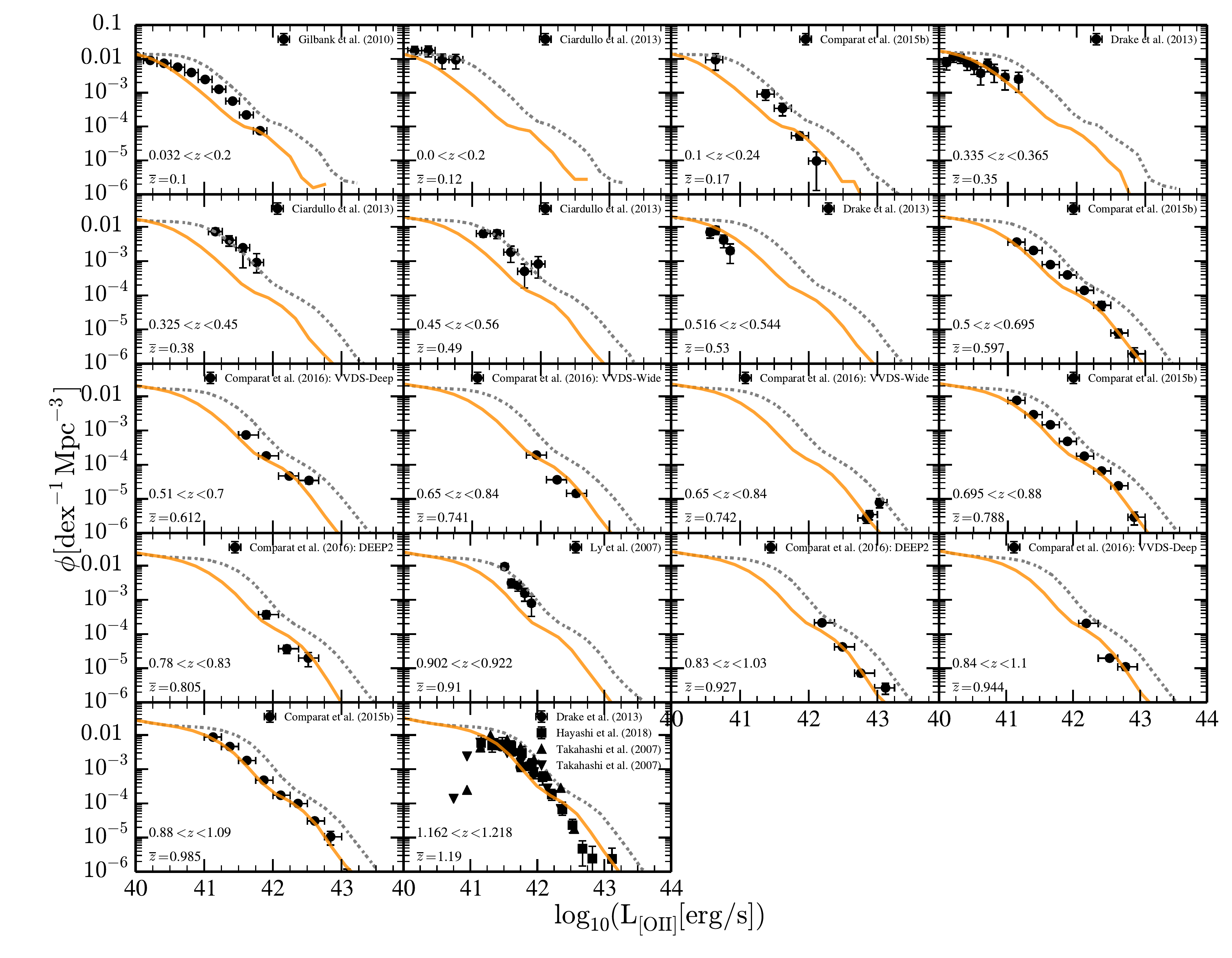}
        \caption{Luminosity function of {\oii} line at 18 different redshifts. Shown are the  comparisons with the observational works of \protect{\cite{Ly2007}},\protect{\cite{Takahashi2007}},\protect{\cite{Gilbank2010}},\protect{\cite{Drake2013}},\protect{\cite{Ciardullo2013}}, \protect{\cite{Comparat2016}}and \protect{\cite{Hayashi2018}}. The black dots represent the observational data, while the solid orange line and grey dashed line the predictions of our mocks LF with and without dust attenuation.}
        \label{fig:ELG_LF_OII}
\end{figure*}

In this appendix we extend  Section~\ref{subsec:linecounts} presenting all the {\Ha}, {\Hb}, {\oii}, and {\oiiiFd} luminosity functions (LF) predicted by our mocks at different redshifts. In Fig.~\ref{fig:ELG_LF_Halpha}, Fig.~\ref{fig:ELG_LF_Hb}, Fig.~\ref{fig:ELG_LF_OIII}, Fig.~\ref{fig:ELG_LF_OII} are presented the LF of the {\Ha}, {\Hb}, {\oiiiFd},  and {\oii} lines, respectively. In all of them black dots represent the observational data, while the solid orange lines and grey dashed lines the predictions of our mock LFs with and without dust attenuation. 

        \
        
\end{document}